\documentclass[manuscript]{aastex}

\begin{document}

\title{The VLA survey of the Chandra Deep Field South III: X--ray
spectral properties of radio sources}


\author{P. Tozzi\altaffilmark{1}, V. Mainieri\altaffilmark{2},
  P. Rosati\altaffilmark{2}, P. Padovani\altaffilmark{2},
  K. I. Kellermann\altaffilmark{3}, E. Fomalont\altaffilmark{3},
  N. Miller\altaffilmark{3,4}, P. Shaver\altaffilmark{2},
  J. Bergeron\altaffilmark{5}, W.N. Brandt\altaffilmark{6},
  M. Brusa\altaffilmark{7}, R. Giacconi\altaffilmark{4},
  G. Hasinger\altaffilmark{7}, B.D. Lehmer\altaffilmark{8},
  M. Nonino\altaffilmark{1}, C. Norman\altaffilmark{4},
  and J. Silverman\altaffilmark{7,9}}

\altaffiltext{1}{INAF Osservatorio Astronomico di Trieste, via
  G.B. Tiepolo 11, I--34143, Trieste, Italy}

 \altaffiltext{2}{ESO, Karl-Schwarzschild-Strasse 2, D-85748 Garching,
   Germany}

\altaffiltext{3}{National Radio Astronomy Observatory, 520 Edgemont
Road, Charlottesville, VA, 22903--2475, USA}

\altaffiltext{4}{Dept. of Physics and Astronomy, The Johns Hopkins
University, Baltimore, MD 21218, USA}

\altaffiltext{5}{Institut d'Astrophysique de Paris, 98bis Boulevard,
  F-75014 Paris, France}

\altaffiltext{6}{Department of Astronomy \& Astrophysics, 525 Davey
  Lab., The Pennsylvania State University, University Park, PA 16802,
  USA}

\altaffiltext{7}{Max--Planck--Institut f\"ur extraterrestrische
  Physik, Giessenbachstra\ss e, PF 1312, 85741 Garching, Germany}

\altaffiltext{8}{Department of Physics, University of Durham, Durham,
  DH1 3LE, UK}

\altaffiltext{9}{Institute of Astronomy, Department of Physics,
  Eidgen\"ossische Technische Hochschule, ETH Zurich, CH8093,
  Switzerland}


\begin{abstract}

We discuss the X--ray properties of the radio sources detected in a
deep 1.4 and 5 GHz VLA Radio survey of the Extended Chandra Deep Field
South (E-CDFS).  Among the $266$ radio sources detected, we find 89
sources (1/3 of the total) with X--ray counterparts in the catalog of
the 1Ms exposure of the central 0.08 deg$^2$ (Giacconi et al. 2002;
Alexander et al. 2003) or in the catalog of the 250 ks exposure of the
0.3 deg$^2$ E-CDFS field (Lehmer et al. 2005).  For 76 (85\%) of these
sources we have spectroscopic or photometric redshifts, and therefore
we are able to derive their intrinsic properties from X--ray spectral
analysis, namely intrinsic absorption and total X--ray luminosities.
We find that the population of submillijansky radio sources with
X--ray counterparts is composed of a mix of roughly 1/3 star forming
galaxies and 2/3 AGN.

The distribution of intrinsic absorption among X--ray detected radio
sources is different from that of the X--ray selected sample.  Namely,
the fraction of low absorption sources is at least two times larger
than that of X--ray selected sources in the CDFS.  This is mostly due
to the larger fraction of star forming galaxies present among the
X-ray detected radio sources.  If we investigate the distribution of
intrinsic absorption among sources with $L_X > 10^{42}$ erg s$^{-1}$
in the hard 2--10 keV band (therefore in the AGN luminosity regime),
we find agreement between the X--ray population with and without radio
emission.  In general, radio detected X--ray AGN are not more heavily
obscured than the non radio detected AGN.  This argues against the use
of radio surveys as an efficient way to search for the missing
population of strongly absorbed AGN.

For the radio sources without cataloged X--ray counterparts, we
measure their average photometric properties in the X--ray bands with
stacking techniques.  We detect emission with very high confidence
level in the soft band and marginally in the hard band.  Given their
redshift distribution, the average X--ray luminosity of these sources
is consistent with being powered by star formation.  We note that on
average, the spectral shape of our radio sources is soft with $HR\sim
-0.5$ and constant in different bins of radio flux.  This results
shows that the statistics do not indicate a significant trend in the
average X--ray spectral properties, but it is consistent with the
radio source population being dominated by star forming galaxies below
$100$ $\mu$Jy, as shown by our morphological and multiwavelength
analysis presented in Mainieri et al. (2008) and Padovani et
al. (2009).
\end{abstract}


\section{Introduction}

Among the most fundamental issues in astrophysics are when and how
galaxies formed and how they evolved with cosmic time.  In particular,
it is crucial to understand the relation between the star formation
processes and the mass accretion history onto the central supermassive
black holes in elliptical galaxies and the bulges of spirals, as
traced by the tight relation between the mass (or the velocity
dispersion) of the bulges and the mass of the supermassive black holes
associated with the Active Galactic Nucleus phase (Kormendy \&
Richstone 1995; Magorrian et al. 1998).  Since these processes have
different signatures throughout the electromagnetic spectrum,
multiband observations are needed to unravel the complex history.

In particular, deep multiwavelength surveys help to reconstruct the
cosmic evolution of AGN and star formation processes.  In this
respect, X--ray and radio emission are good tracers of both processes.
The radio properties of the X--ray population found in deep surveys
have been studied in a few papers based on VLA data in the
Chandra Deep Field North (CDFN; Richards et al. 1998; Richards 2000;
Bauer et al. 2002; Barger et al. 2007), combined MERLIN and VLA data
in the CDFN region (Muxlow et al. 2005), and ATCA data in the Chandra
Deep Field South (CDFS; Afonso et al. 2006; Rovilos et al. 2007).
Deep radio surveys are also realized in shallower but wider X--ray
fields like COSMOS (see Schinnerer et al. 2007; Smolcic et al. 2008a;
2008b).  In this Paper, we use the deep radio data obtained with the
VLA in the CDFS and Extended Chandra Deep Field South (E-CDFS) fields.
The comparison of the properties of the radio sources (whose catalog
is presented in Kellermann et al. 2008, hereafter Paper I) and of the
X--ray sources (see Giacconi et al. 2002; Alexander et al. 2003;
Lehmer et al. 2005) allows us to characterize both processes over a
wide range of redshifts.

In this Paper we present a systematic study of the X--ray properties
of the radio sources in the CDFS radio catalog.  The radio catalog
includes 266 sources (see Paper I) and constitutes one of the largest
and most complete samples of $\mu$Jy sources in terms of redshift
information.  We have redshifts for 186 ($\sim 70$\%) of the sources,
108 spectroscopic and 78 photometric.  We have reliable
optical/near--IR identifications for 94\% of the radio sources, and
optical morphological classifications for $\sim$ 61\% of the sample.
Optical and near--IR properties of the radio sources are discussed by
Mainieri et al. (2008, hereafter Paper II), while a multiwavelength
approach to studying the source population is presented by Padovani et
al. (2009, Paper IV).

The present Paper is organized as follows.  In \S 2 we briefly
describe the radio, X--ray and optical data sets.  In \S 3 we describe
the procedure used to identify the X--ray counterparts of the radio
sources.  In \S 4 we describe the X--ray properties of the radio
sources with X--ray counterparts in the catalog of Giacconi et
al. (2002) for the CDFS and of Lehmer et al. (2005) for the E-CDFS.
In \S 5 we show the average X--ray properties of radio sources without
individual X--ray counterparts, obtained by stacking techniques.  Our
conclusions are summarized in \S 6.  Luminosities are quoted for a
flat cosmology with $\Omega_{\Lambda}=0.7$ and $H_0=70$ km/s/Mpc (see
Spergel et al. 2006).

\section{The data}

\subsection{The Radio data}

We observed the whole area of the E-CDFS ($\sim 0.3$ deg$^2$) with the
NRAO Very Large Array (VLA) for 50 h at 1.4 GHz mostly in the BnA
configuration in 1999 and February 2001, and for 32 h at 5 GHz mostly
in the C and CnB configurations in 2001.  The effective angular
resolution is 3.5'' and the minimum {\sl rms} noise is as low as 8.5
$\mu$Jy per beam at both 1.4 GHz and 5 GHz. These deep radio
observations complement the larger area, but less sensitive ({\sl rms}
$\sim 14 \mu$Jy per beam), lower resolution observations of the CDFS
discussed by Afonso et al. (2006)

Here we use the radio catalog presented in Paper~I.  A total of 266
radio sources were catalogued at 1.4 GHz, 198 of which are in a
complete sample with signal--to--noise ratio greater than 5, and
located within 15' from the field center.  The corresponding flux
density limit ranges from 42 $\mu$Jy at the field center, to 125
$\mu$Jy near the field edge.  Further discussion of the radio sources
found in a larger area survey which includes the full E-CDFS with a
uniform rms noise level of $\sim 6 \mu$Jy (Miller et al. 2009) will be
given in a later paper.

The catalog includes radio positions, 1.4 GHz and 5 GHz flux
densities, signal--to--noise ratios at the two frequencies, the
largest angular size, and the radio spectral index between 5 and 1.4
GHz.  Flux densities at 5 GHz are available only for $\sim$ 70\% of
the sources (187 out of 266).

Among these sources, 22 have multiple components (12 are double, while
8 have three components, and only 2 sources have four components).
The multiple component sources are associated mainly with classical
radio galaxies.  At least half of the components of multiple sources
have a maximum extension larger than 3'', while only 1/3 of the
single--component sources have extension larger than 3''.  Clearly,
the classification of a source as compact or extended depends on the
spatial resolution of the radio data.  Here we treat all the 266
sources in the catalog as single, and use the centroid for the
multiple sources.  The secondary components of multiple sources (those
components which do not correspond to the centroid, but most likely to
a radio lobe), are treated separately.

\subsection{The X--ray data}

In the E-CDFS area, we have two sets of X--ray data.  The first is the
1Ms exposure in the central $\sim 0.1$ deg$^2$ (Rosati et al. 2002;
Giacconi et al. 2002; Alexander et al. 2003), the second one is the
shallower $\sim 250$ ks coverage of a square region of $0.28$ deg$^2$
centered on the above field (Lehmer et al. 2005).  The two data sets
are treated separately, since it is not convenient to add them due to
the large differences in the point spread function in the overlapping
areas.  Therefore, we use the deeper 1Ms data whenever the effective
exposure time is larger than 25\% of the effective exposure at the
aimpoint (940 ks), while we use the shallower and wider E-CDFS data in
the remaining area, where the quality of the E-CDFS exposure is better
than the CDFS one.  In this way we avoid regions close to the border
of the 1Ms image, where the low effective exposure and the broadening
of the PSF make the quality of data lower than that of the E-CDFS in
the same region.

The 1Ms dataset of the CDFS is the result of the coaddition of 11
individual {\it Chandra} ACIS--I (Garmire et al. 1992; Bautz et
al. 1998) exposures with aimpoints spaced within a few arcsec from
$\alpha=$3:32:28.0, $\delta=-$27:48:30 (J2000).  For the X--ray data
reduction of the CDFS--1Ms sources, we used the software {\tt ciao}
3.0.1\footnote{For the latest version see
  http://cxc.harvard.edu/ciao3.4/index.html} and the calibration
database CALDB 2.26\footnote{For the release notes see
  http://asc.harvard.edu/caldb/downloads/Release\_notes/CALDB\_v2.26.txt},
therefore including the correction for the degraded effective area of
ACIS--I chips due to material accumulated on the ACIS optical blocking
filter at the epoch of the observation.  We also apply the
time--dependent gain correction\footnote{see
  http://asc.harvard.edu/ciao/threads/acistimegain/}.  The reduction
and analysis of the X--ray data are described in more detail in
Giacconi et al. (2001), Tozzi et al. (2001) and Rosati et al. (2002).
The final image covers 0.108 deg$^2$, where 347 X--ray sources are
identified down to flux limits of $5.5 \times 10^{-17}$ and $4.5\times
10^{-16}$ erg cm$^{-2}$ s$^{-1}$ in the soft ($0.5-2$ keV) and hard
($2-10$ keV) bands respectively.  In this Paper we will refer to the
X--ray catalog presented in Giacconi et al. (2002).

X--ray spectral properties of the sources in the Giacconi et al.(2002)
catalog have been presented in Tozzi et al. (2006).  However, new
redshifts have since been found for a significant number of sources,
as a result of the ongoing spectroscopic follow up of the X--ray and
radio sources in the E-CDFS field.  In particular, 19 X--ray sources
with radio counterparts in the CDFS field have new or updated
redshifts with respect to Szokoly et al. (2004), while 27 X--ray
sources with radio counterparts in the E-CDFS have new unpublished
redshifts.  For these sources the X--ray spectral analysis is updated
consistently.  Spectra are fitted with a power law (XSPEC model {\tt
  pow}\footnote{See
  http://heasarc.nasa.gov/xanadu/xspec/manual/XSmodelPowerlaw.html})
with intrinsic absorption at the source redshift (XSPEC model {\tt
  zwabs}\footnote{http://heasarc.nasa.gov/xanadu/xspec/manual/XSmodelWabs.html})
with redshift fixed to the spectroscopic or photometric value.  We
also include a redshifted K--shell Fe line modeled as an unresolved
Gaussian component at $6.4/(1+z)$ keV (Nandra \& Pounds 1994).  We
take into account the local Galactic absorption (XSPEC model {\tt
  tbabs}\footnote{See
  http://heasarc.nasa.gov/xanadu/xspec/manual/XSmodelTbabs.html}) with
a column density frozen to $N_H = 8 \times 10^{19}$ cm$^{-2}$ (from
Dickey \& Lockman 1990).  In performing the spectral fits, we include
the effects of a methylen layer which is not yet accounted for in the
calibration release CALDB 2.26 (see Vikhlinin et al. 2005).  We use
XSPEC v11.3.1 (see Arnaud 1996) to perform the spectral fits.

The E-CDFS survey consists of four contiguous $\sim 250$ ks Chandra
observations covering approximately $\simeq 0.3$ deg$^2$, flanking the
1Ms CDFS.  The data and the point--source catalog are presented in
Lehmer et al. (2005).  The survey reaches flux limits of $1.1 \times
10^{-16}$ and $6.7 \times 10^{-16}$ erg cm$^{-2}$ s$^{-1}$ in the
$0.5-2$ keV and $2-8$ keV bands, respectively, and it includes 755
point--sources, of which 583 are not previously detected in the 1Ms
exposure of the CDFS, mostly because of the larger covered area.  For
the X--ray data reduction of the E-CDFS sources, we used {\tt ciao}
3.1 and CALDB 2.29.  The X--ray spectral analysis of radio sources
with counterparts present only in the E-CDFS catalog relies on new
redshifts obtained during the spectroscopic follow--up of the E-CDFS
and are presented here for the first time.  The spectral analysis
procedure is the same as that used for the sources identified in the
1Ms data.

\subsection{The optical data}

For the sources identified in the 1Ms exposure of the CDFS, the
spectroscopic identification program carried out with the ESO--VLT is
presented in Szokoly et al. (2004).  The optical classification is
based on the detection of high ionization emission lines.  The
presence of broad emission lines (width larger than $2000$ km/s) like
Mg$II$, C$III$, and at large redshifts, C$IV$ and Ly$\alpha$,
classifies the source as a Broad Line AGN (BLAGN), Type--1 AGN or QSO
according to the simple unification model by Antonucci (1993). The
presence of unresolved high ionization emission lines (like O$III$,
Ne$V$, Ne$III$ or He$II$) classifies the source as a High Excitation
line galaxy (HEX), often implying an optical Type--2 AGN
classification.  Objects with unresolved emission lines consistent
with an H$II$ region spectrum are classified as Low Excitation Line
galaxies (LEX), implying sources without optical signs of nuclear
activity.  However, discriminating between a Type-2 AGN and an $H II$
region galaxy involves the measurement of line ratios as shown in
Veilleux \& Osterbrock (1987), which is not used here as a
classification scheme.  Objects with typical galaxy spectra showing
only absorption lines are classified as ABS.  Among the LEX class we
expect to find star--forming galaxies or Narrow Line Emission
Galaxies, but also hidden AGN.  Hidden AGN may be present also in the
ABS class.  The optical identification is flagged according to the
quality of the optical information.  In several cases, the optical
spectral properties do not allow us to obtain a secure determination
of the spectral type.  As shown in Szokoly et al. (2004), the optical
classification scheme fails to identify as AGN about 40\% of the
X--ray sources in the LEX+ABS classes.  Therefore, an X--ray
classification scheme, based on the source hardness ratio and observed
X--ray luminosity, was developed by Szokoly et al. (2004) and compared
with the optical classification (see their Fig.~13).  A refined
X--ray/optical classification scheme, based on X--ray spectral
analysis, is presented in Tozzi et al. (2006).  Optical and near-IR
images of the CDFS are also used to derive photometric redshifts for
all the X--ray sources without spectroscopic data.  Using the widest
multiwavelength photometry available today, Zheng et al. (2004) and
Mainieri et al. (2005) derived photometric redshifts for the entire
sample of optically identified CDFS X--ray sources.

Radio sources are identified with optical conterparts in Paper II.  In
some cases, the radio data and the use of new optical and MIR data
from {\sl Spitzer} allowed us to better identify the counterpart of
some of the X--ray sources (see Brusa et al. 2008, in preparation).
In these cases the X--ray spectral analysis of the source is updated
with the new redshift resulting from the new identification, as
discussed above.  For several radio sources, the optical spectra of
the counterparts are obtained in the follow--up of E-CDFS sources (see
Silverman et al. 2008).

\section{Matching radio sources with X--ray data}

The radio catalog is presented in Paper I.  To investigate the X--ray
properties of the radio sources, first we match the radio sources with
the X--ray catalogs of Giacconi et al. (2002, after applying the
posititional shift correction as in Alexander et al. 2003), whenever
their exposure time in the X--ray image is larger than 25\% of the
maximum exposure of the CDFS--1Ms field.  For all the remaining radio
sources, we match them with the catalog of Lehmer et al. (2005) in the
E-CDFS field. To identify X--ray counterpart candidates, we initially
selected all the pairs of radio and X--ray sources with separation
less than $3 \sigma_d$, where $\sigma_d^2= \sigma_r^2 + \sigma_X^2$
and $\sigma_R$ and $\sigma_X$ are the rms error of the radio and
X--ray positions respectively.  Typically, $\sigma_x$ ranges from
0.2'' to 1.5'' depending on the off--axis angle, as computed by
Giacconi et al. (2002), while $\sigma_r$ ranges from 0.5'' to 2''.
\footnote{Radio positions are listed in Paper I.}.  If more
than one source satisfies this criterion, the preferred counterpart is
the one with the smallest offset.  However, this criterion for
radio/X--ray source matching was refined as described below.

First we examined the 126 radio sources in the CDFS-1Ms observation.
Following the matching procedure discussed above, we searched for
X--ray counterparts and find 55 matching candidates.  Then, we use the
optical identifications both of the radio and the X-ray sources (see
Paper II) to refine the positions and check for possible false
matches.  We then produced thumbnails of the radio, X--ray and optical
images for all the candidates.  Optical images are chosen from the
available bands, using a priority based on depth and spatial
resolution: R--band with FORS at VLT (see Giacconi et al. 2002),
z--band GEMS (Rix et al. 2004), i--band ACS for GOODS (Giavalisco et
al. 2004), WFI R deep (Hildebrandt et al. 2006).  A close visual
inspection, with the help of the optical images which have the best
resolution, allowed us to discard three likely false matches, whose
X--ray counterpart candidates are associated with optical sources
different from the optical counterparts of the radio sources (see the
contour maps of the extended sources overlaid over the WFI images with
the position of the X--ray sources shown in Paper~I).  We also
visually investigated the radio sources without X--ray match
candidates to look for missed matches, but found none.  The X--ray
counterparts from the Giacconi and Lehmer catalogs are shown in
Table I of Paper I

Down to the CDFS--1Ms flux limits ($5.5\times 10^{-17}$ ergs cm$^{-2}$
s$^{-1}$ in the soft 0.5--2 keV band and $4.5 \times 10^{-16}$ ergs
cm$^{-2}$ s$^{-1}$ in the hard 2-10 keV band) we have 52 radio sources
with X--ray counterparts (corresponding to 40\% of the whole radio
sample) and 74 radio sources without X--ray counterparts.

For the remaining radio sources we searched for X--ray counterparts in
the E--CDFS data with the same procedure.  We find 37 (corresponding
to 26\%) X--ray matches (after removing one false match).  In total,
we thus have 89 radio sources with X--ray counterparts (52 from the
CDFS-1Ms data, and 37 from the E-CDFS data).  This number is exactly
the same found by Rovilos et al. (2007), where the ATCA data by Afonso
et al. (2006) are used.  However, we compared our X--ray detected
radio sources with those of Rovilos et al. (2007), and we found
significant differences: we have 31 X--ray detected radio sources not
included in Rovilos et al, while Rovilos et al. includes 31 sources
which we do not include.  Among them, 23 are radio sources not present
in our catalog, one has been discarded as a false match, and 7 are not
included because of our matching criterion.  The main difference in
the two samples is accounted by the differences of the two radio
surveys, both in sensitivity (Kellerman et al. 2008 is more sensitive
in the center, while Afonso et al. 2007 in the outer regions) and in
the accuracy of the radio positions.

The majority of radio sources (177) do not have X--ray counterparts in
the X--ray catalogs, but were studied with stacking techniques.  In
summary, about 1/3 of the radio sources cataloged in Paper I have an
X--ray counterpart in the present analysis.


\section{Properties of radio sources with X--ray counterparts}

For all of the radio sources with an X--ray counterpart and redshift
information, we analyze the X--ray spectrum as described in Tozzi et
al. (2006).  We have 89 radio sources with a cataloged X--ray
counterpart, 76 of them with known redshifts.  Among them, 31 have
spectroscopic redshifts and optical classification based on the
detection of optical emission lines, while the remaining 45 sources
have only photometric redshifts or uncertain optical classification.

\begin{figure}
\plottwo{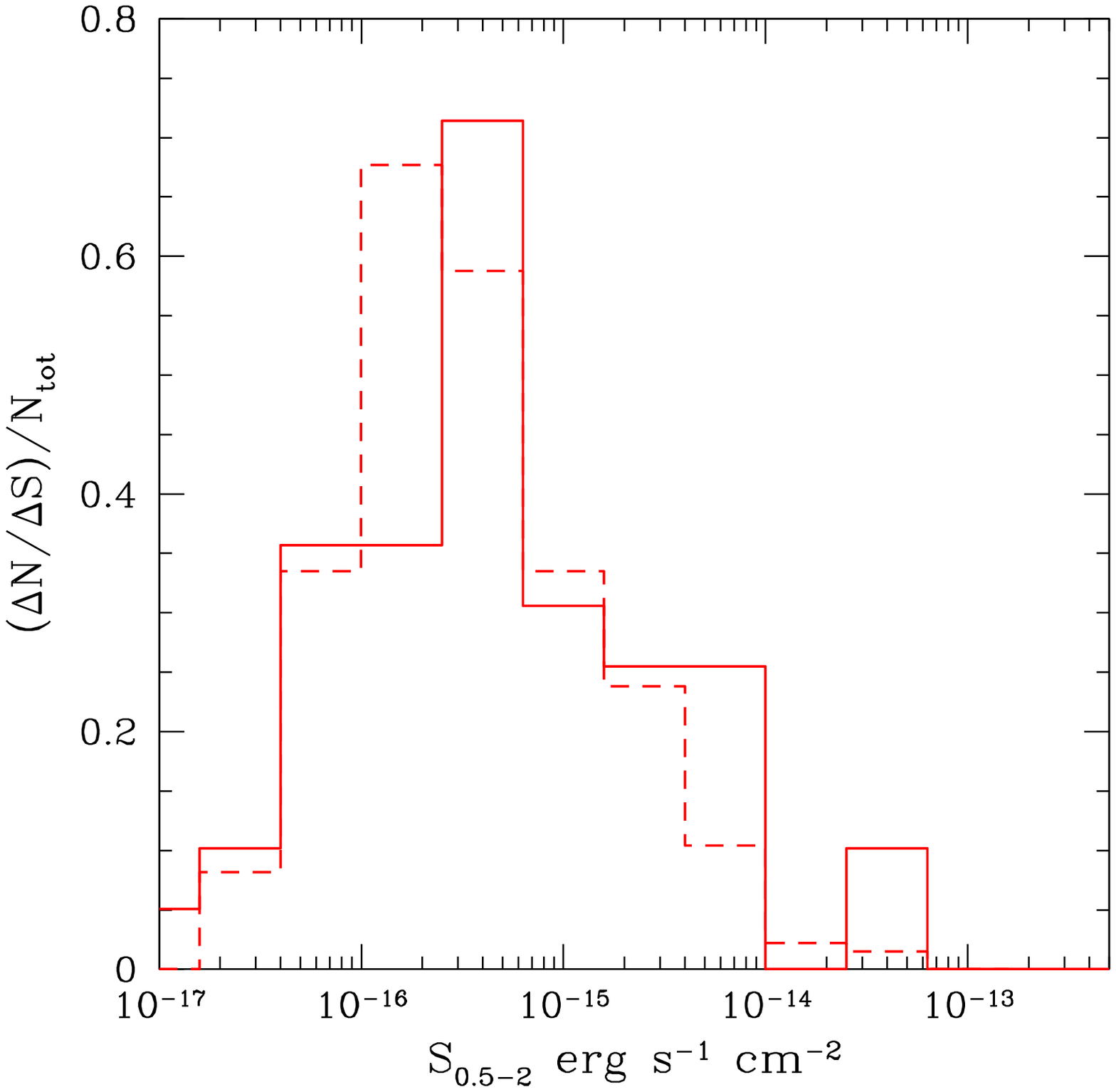}{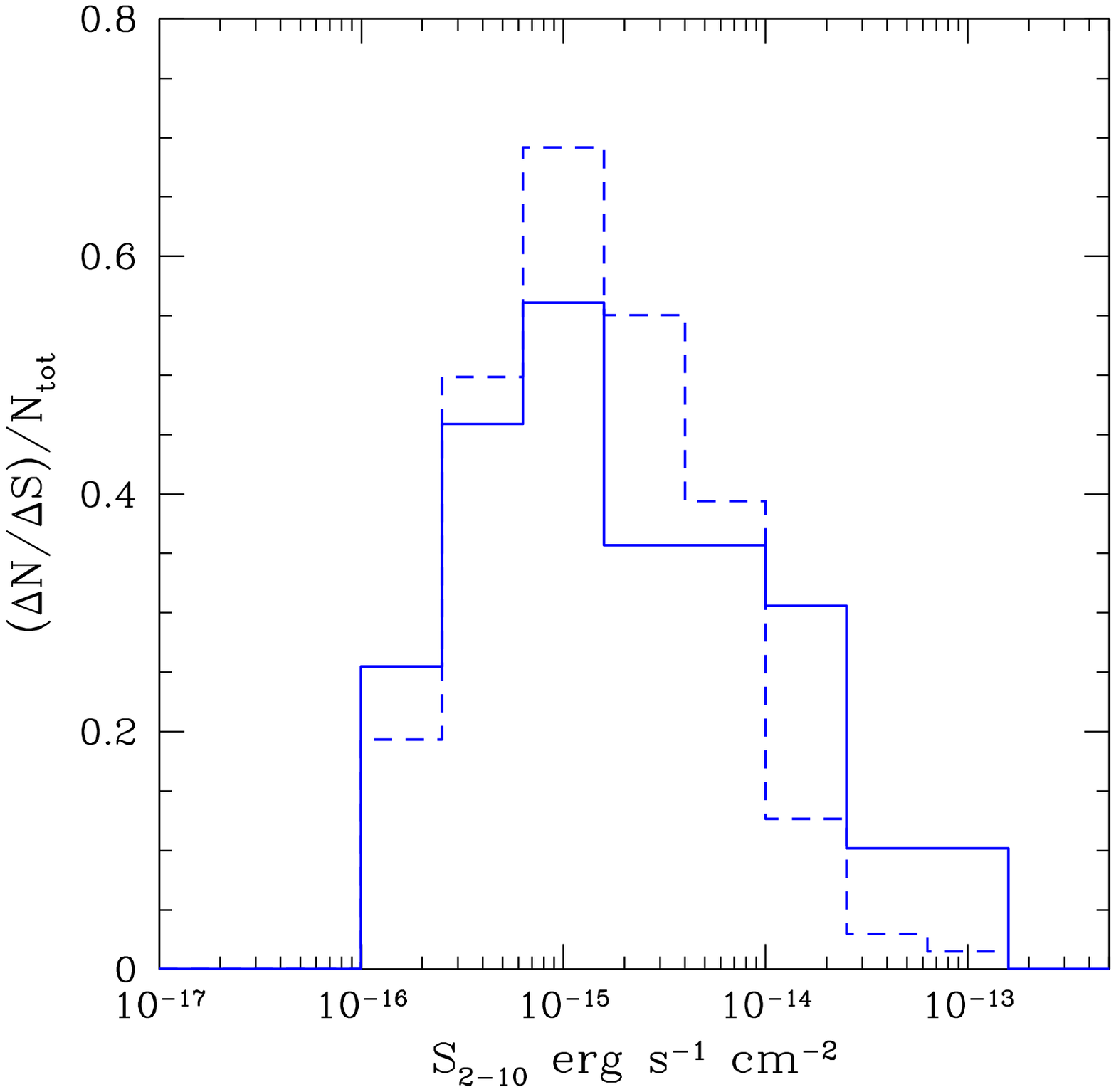}
\caption{{\sl Left panel}: normalized distribution of X--ray fluxes in
  the soft band for all the radio sources with X--ray counterparts in
  the 1 Ms CDFS field (solid line), compared with that of the whole
  X--ray sample (dashed line).  {\sl Right panel:} normalized
  distribution of X--ray fluxes in the hard band for all the radio
  sources with X--ray counterparts in the 1Ms field (solid line),
  compared with that of the whole 1 Ms CDFS X-ray sample (dashed
  line). }
\label{flux_distrib}
\end{figure}

\begin{figure}
\plotone{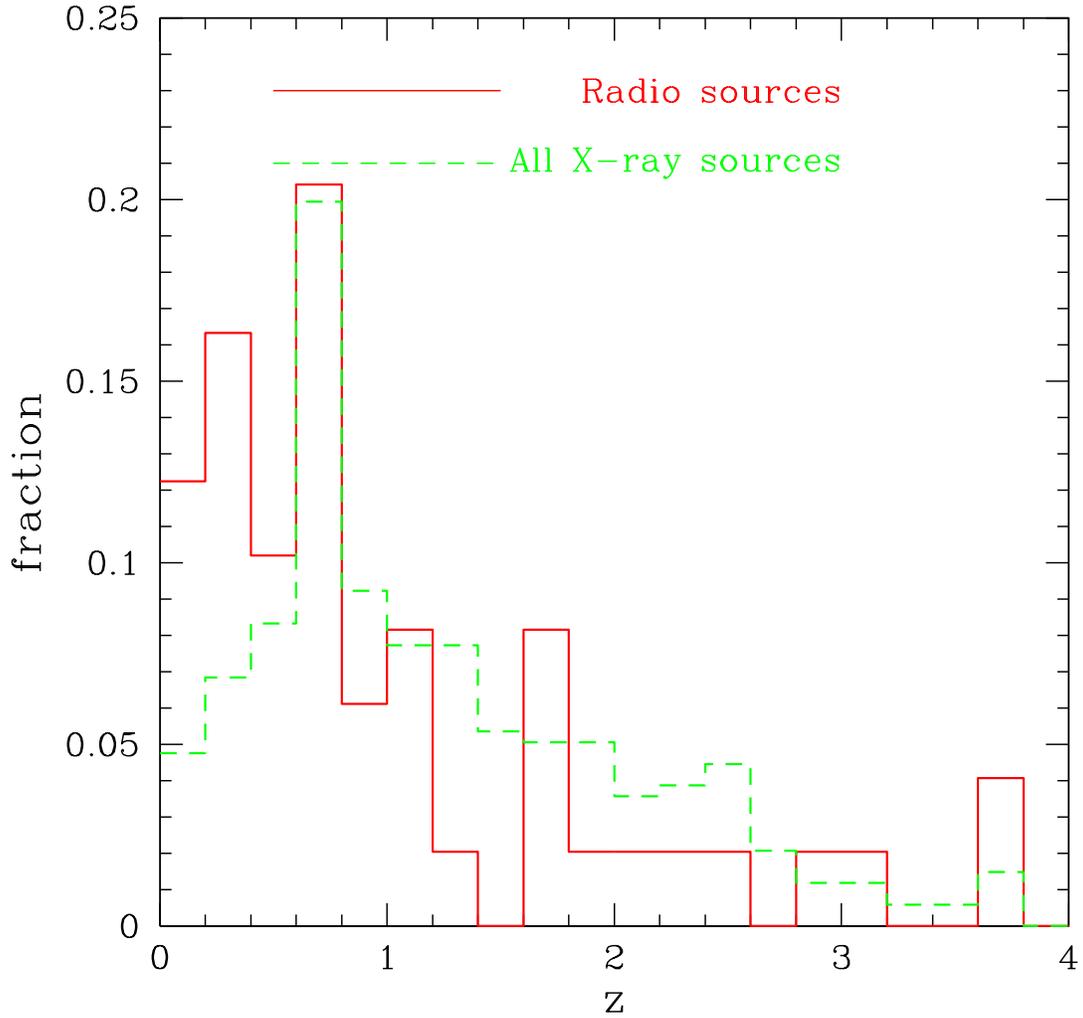}
\caption{Redshift distribution of the 49 radio sources with 1 Ms CDFS
  X--ray counterparts and redshift information compared with the
  redshift distribution of the X--ray sources in the Giacconi et
  al. (2002) catalog.  }
\label{zradio}
\end{figure}

The normalized distribution of X--ray fluxes of the radio sources with
X--ray counterparts found in the 1Ms field is shown in Figure
\ref{flux_distrib} for the soft (left panel) and hard (right panel)
bands.  The distributions are compared with those of the whole X--ray
sample in the 1Ms exposure of the CDFS.  We find that the radio
selection at the current flux density limit marginally tends to
consist of the brightest X--ray sources.  From a KS test, we find that
the probability of the two distributions being extracted from
different parent populations are $\sim 87$\% and $\sim 95$\% for the
soft and hard bands, respectively.  None of these probabilities are
significant (i.e., $> 95$\%), therefore we conclude that the
additional brightness limit introduced by the radio selection is not
affecting much the X--ray flux distribution.  On the other hand, the
normalized redshift distribution of the radio sources with X--ray
counterparts is significantly shifted towards lower redshifts with
respect to the distribution of the whole X--ray sample, as shown in
Figure \ref{zradio}.  The average redshift of the radio sources is
$\langle z \rangle = 1.01$ (median 0.73) while that of the X--ray
sources is $\langle z \rangle = 1.28$ (median 1.03).  The two redshift
distributions are inconsistent at more than the 99\% confidence level.
This is mostly due to the larger fraction of star forming galaxies
among the X--ray sources with radio counterpart, since the radio
emission is often associated with star formation.  Since star forming
galaxies are intrinsically fainter in the X--ray band, they are
typically found at lower redshift with respect to AGN among the CDFS
X-ray sources.  This is the main reason of the shift towards lower
redshift among radio sources with X--ray counterparts.  Note that the
peak at $z\sim 0.7$ is due to the large-scale structure noted in Gilli
et al. (2003).

Most of the sources with spectroscopic redshifts have also an
unambiguous optical classification based on the detection of high
ionization emission lines, as described in \S 2.2.
However, only 31 sources in the Szokoly et al. (2004) sample have
spectroscopic redshifts and optical classification based on the
detection of optical emission lines.  We notice that among the radio
sources with X--ray counterparts, a wide range of optical types are
present.  We find 12 radio sources distributed among BLAGN (5) and HEX
(7), which corresponds roughly to Type I and Type II AGN respectively
(see Szokoly et al. 2004; Tozzi et al. 2006).  The most common optical
species corresponds to LEX, which includes 14 radio sources.  Among
these sources we expect a larger number of star forming galaxies or
Narrow Line Emission Galaxies, but also hidden AGN.  Only 5 sources
are in the ABS spectroscopic class.

\begin{figure}
\plotone{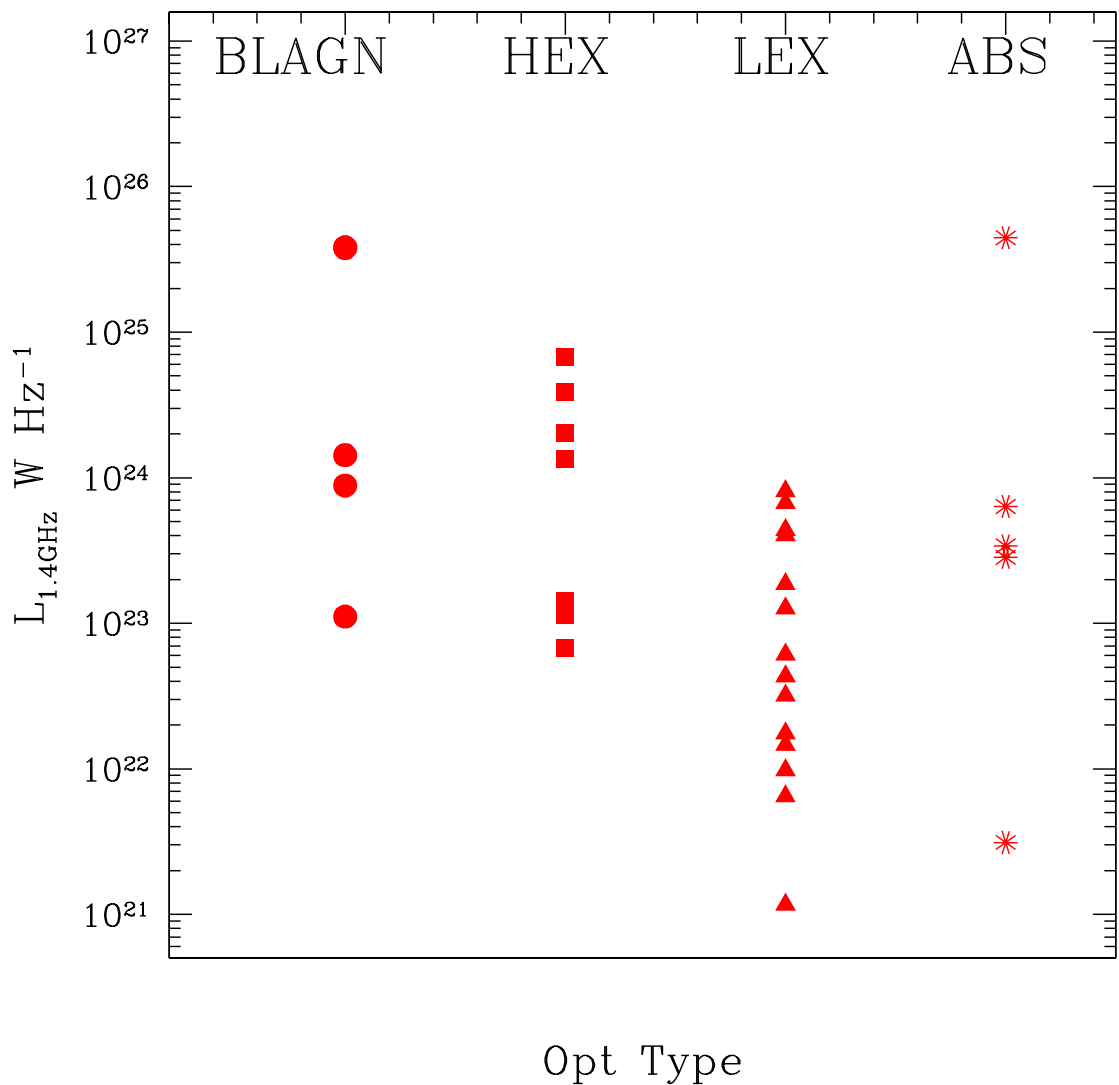}
\caption{Radio luminosities plotted for different optical classes:
  Broad Line AGN (circles) High Excitation Lines (Squares), Low
  Excitation Lines (triangles) and normal galaxies (asterisks).
  K--corrections are computed for the measured $\alpha_R$ when
  possible, otherwise assuming $\alpha_R=0.7$.  }
\label{optype}
\end{figure}

The distribution of radio luminosities is different for the four
optical types, as shown in Figure \ref{optype}.  The radio luminosity
density is computed using the measured radio spectral slope when
available (see Paper I), while we assume the average value $\alpha _R=
0.7$ for sources detected only at 20 cm.  The rest--frame radio
luminosity density was calculated as $L_{1.4GHz} = 4\pi d_L^2
S_{1.4GHz} 10^{-33} (1+z)^{\alpha_R-1}$ W Hz$^{-1}$ where $d_L$ is the
luminosity distance (cm) and $S_{1.4GHz}$ is the flux density ($mJy$).
The radio luminosity of BLAGN ranges from $ 10^{23}$ to a few $\times
10^{26}$ W Hz$^{-1}$, while for HEX sources it reaches only $ 10^{25}$
W Hz$^{-1}$.  The LEX radio sources are mostly in the range from
$10^{21}- 10^{24}$ W Hz$^{-1}$, typical of star forming galaxies.
Sources in the ABS optical class have radio luminosities consistent
with Faranoff--Riley Type I galaxies and star-forming galaxies for the
lower luminosity sources, except one bright radio galaxy.
 

All of the information for the 76 radio sources with an X--ray
counterpart and redshift  are shown in Table \ref{table1}
and Table \ref{table2} for the 1Ms and the E-CDFS fields respectively.
X--ray luminosities are obtained from the X--ray spectral analysis,
and refer to the intrinsic (de--absorbed) emitted power.  For 13 radio
sources with X--ray counterpart in the CDFS or in the E-CDFS fields,
we do not have redshift information.  These sources are shown in Table
\ref{table3} and are excluded from the X--ray spectral analysis.

\begin{figure}
\plotone{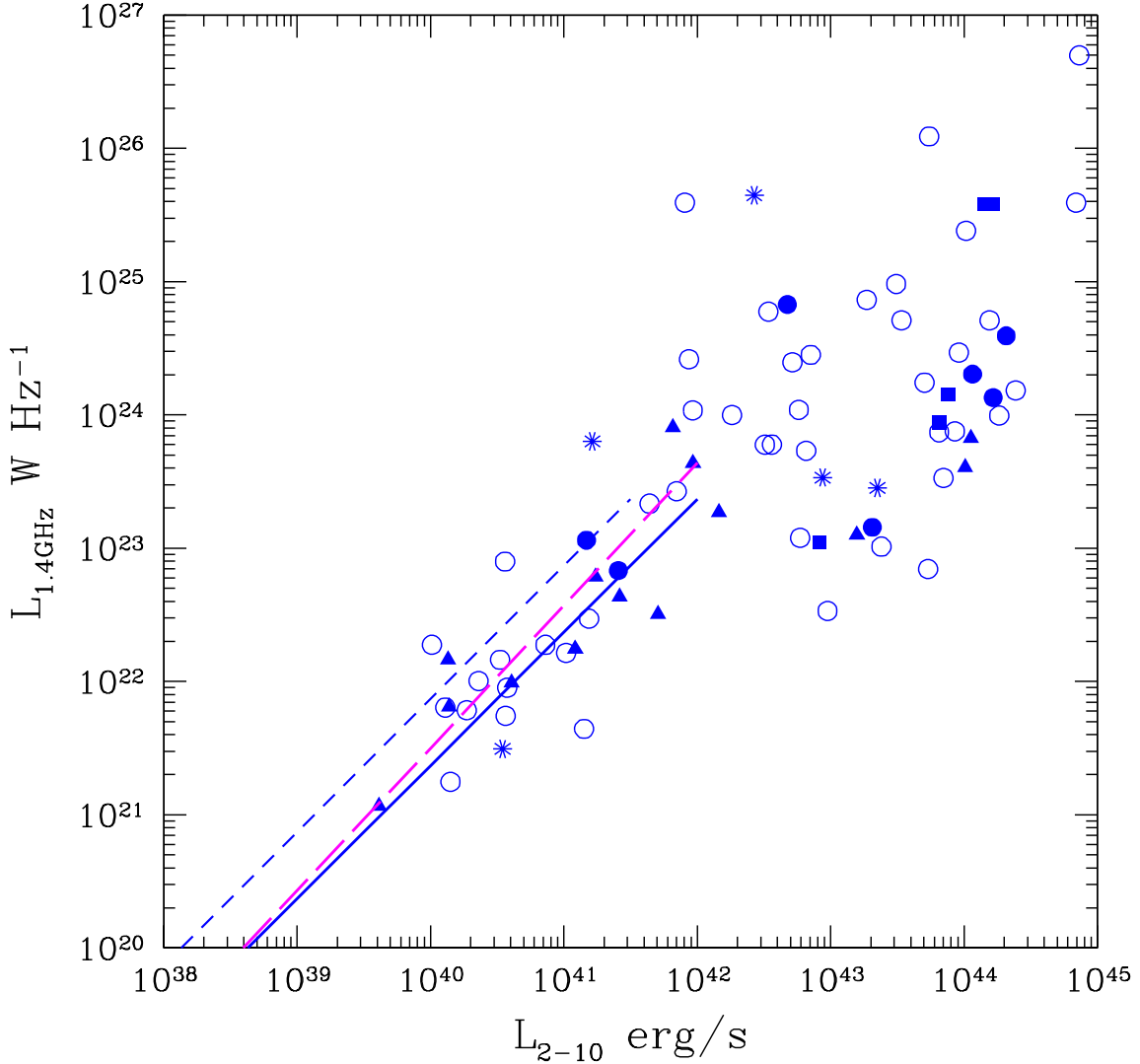}
\caption{Radio luminosity plotted against the X--ray luminosity in the
  hard band (2--10 keV).  Same symbols as in Figure \ref{optype}
  (empty circles for sources without optical spectral classification).
  The solid line is the correlation between radio and hard band X--ray
  luminosity determined empirically for star forming galaxies in
  Persic \& Raphaeli (2007), while the short-dashed line is the same
  relation found by Ranalli et al. (2003), and the long-dashed line is
  by Bauer et al. (2002).  K--corrections are computed for the
  measured $\alpha_R$ when possible, otherwise assuming
  $\alpha_R=0.7$.
}
\label{lum}
\end{figure}

In Figure \ref{lum}, we plot the 20 cm radio and hard--band X--ray
luminosities for the 76 radio sources with X--ray spectral analysis
and redshift information.  We recall here that the X--ray luminosities
represent the intrinsic power emitted in the corresponding rest--frame
X--ray band (after removing intrinsic absorption), and the X--ray
K--correction is already accounted for via the detailed spectral
analysis.  We find a clear trend in the luminosity range typical of
star forming galaxies ($10^{40} \leq L_X \leq 10^{42}$ erg s$^{-1}$)
and a larger scatter for higher X--ray luminosities.  This result also
reinforces the expectation that the majority of sources in our sample
with $L_X \leq 10^{42}$ erg s$^{-1}$ are powered by star forming
activity both in the radio and in the X--ray bands. Indeed, by looking
at the sources with optical classification, the star forming galaxies
sector may be conservatively defined by the conditions $log(L_x)\simeq
41.5$ and $log(L_{1.4}\simeq 23$.  However, this result is based only
on 40\% of the sources and must be taken with caution.  Among 14
sources in the SF-galaxy luminosity range, the dominant optical types
are LEX (10 sources) and ABS (2 sources), with only 2 HEX sources.

We notice also that the correlation expected for star forming
galaxies, as found by Ranalli et al.  (2003) for the X--ray luminosity
range $10^{38} < L_X < 10^{41.5}$ in the hard band, appears to be a
factor $\sim 2$ higher.  On the other hand, the radio-X--ray
luminosity relation by Persic \& Rephaeli (2007) for the integrated
X---ray hard luminosity at high redshift (see their equations 10 and
11) is in better agreement with the X--ray detected radio sources.
The same holds for the $L_R-L_X$ relation found for 122 late type
galaxies in the CDFN studied by Bauer et al. (2002).  A censored
statistical analysis including the X--ray upper limits, confirms the
linear slope and normalization of Ranalli et al. (2003), as shown in
Paper IV.  The presence of a robust $L_X-L_R$ relation agrees also
with the recent analysis by Lehmer et al. (2008) on a sample of late
type galaxies in the CDFN and E-CDFS, suggesting that X--ray emission
can be used as a robust indicator of star formation activity out to
$z\sim 1.4$.  All the mentioned studies, including our own, are at
variance with the argument of Barger et al. (2007) that the
$L_X$--$L_R$ relation is spurious.

On the other hand, the wide scatter at higher luminosities reflects
the wide range of radio to X--ray luminosity ratio found in
AGN. Within the 17 sources with $L_X>10^{42}$ erg s$^{-1}$ , the
dominant optical types are distributed among BLAGN (5 sources), HEX (5
sources), LEX (4 sources) and ABS (3 sources).  While X--ray emission
in this luminosity range is mostly associated to the AGN, the radio
emission may be still associated to star formation activity in the
majority of sources, as found by Rovilos et al. (2007) using the
Spitzer 24 $\mu$m luminosity.

Variability of the X--ray and radio luminosity may increase the
scatter for the AGN sources.  In order to check this possible bias, we
considered the X--ray and radio luminosities separately for the
sources with and without detected variability (Paolillo et al. 2004).
Only the 46 sources for which variability has been measured (in the
1Ms exposure field) are included. We do not find any statistical
evidence for a different behaviour (i.e., a larger scatter) among the
two subsamples, which include 14 and 32 variable and non-variable
sources respectively.  Therefore, it seems unlikely that X--ray
variability can account for a significant part of the large observed
scatter in the $L_X$--$L_R$ relation for AGN.  However, we cannot
completely exclude some effect, since, as shown in Paolillo et
al. (2004), probably the large majority ($>90$\%) of the CDFS sources
are X--ray variable, their variability being undetected due to the low
SNR.

\begin{figure}
\plotone{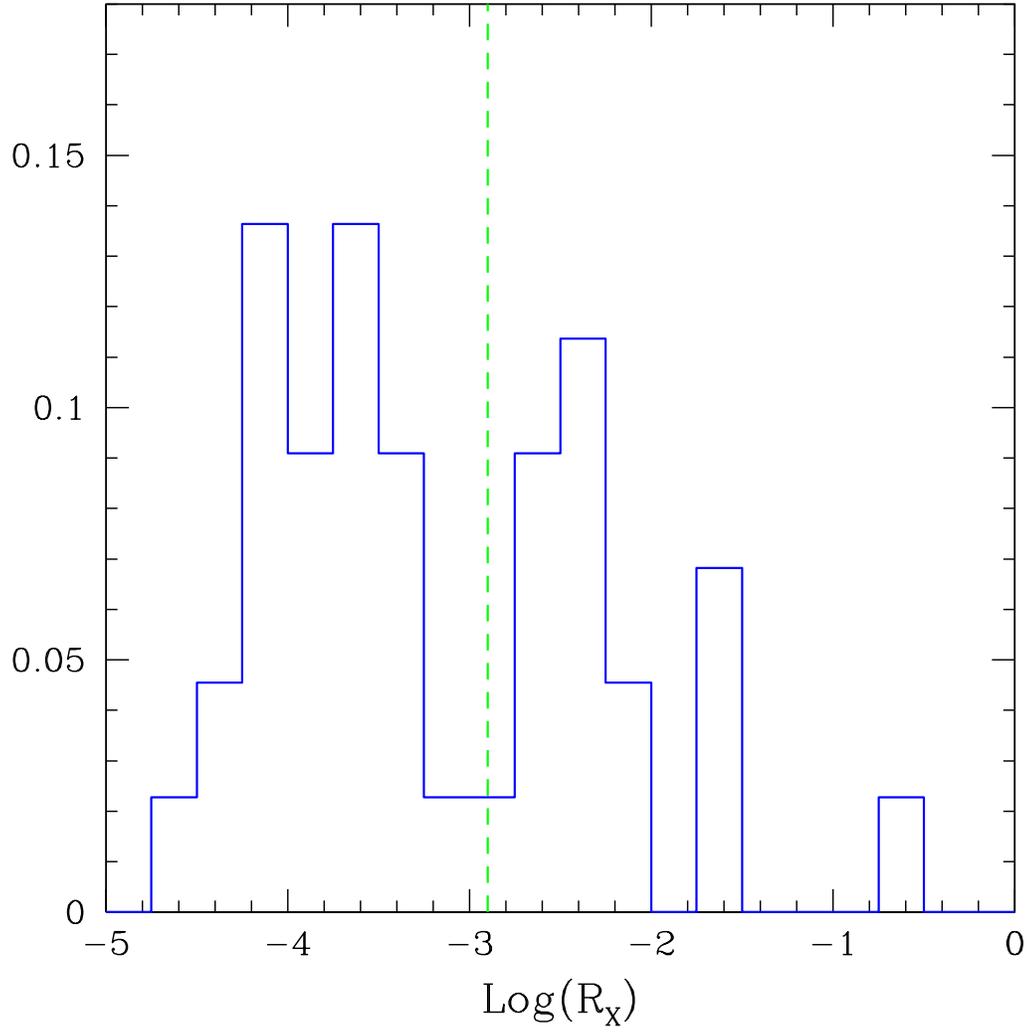}
\caption{Distribution of the radio loudness $\log(R_X)$ for sources
  with $L_{2-10} > 10^{42}$ erg s$^{-1}$.  The vertical dashed line
  shows the boundary between radio loud and radio quiet AGN after
  Panessa et al. (2007).}
\label{rx}
\end{figure}

We also compute the radio loudness with respect to the hard band
X--ray flux as defined in Terashima \& Wilson (2003).  Here we use the
20 cm luminosity rather than the 6 cm luminosity used by Terashima \&
Wilson: $R_X \equiv \nu L_R(5GHz) /L_{2-10}$.  On average, radio
loudness defined by the 20 cm luminosity is shifted by $-0.16$ with
respect to $R_X$ defined at 6cm.  Therefore we take $log(R_X) = -2.9$
as the boundary between radio loud and radio quiet AGN found by
Panessa et al. (2007) with a sample of local Seyfert galaxies and
low--luminosity radio galaxies.  Since this criterion applies only to
sources with nuclear activity, we consider only sources with $L_X >
10^{42}$ erg s$^{-1}$.  As shown in Figure \ref{rx}, the radio
loudness distribution of the whole sample of sources with X--ray
spectral analysis shows some bimodality.  Using the same radio
loud/radio quiet boundary, among the $\sim 50$ sources with $L_{2-10}
> 10^{42}$ erg s$^{-1}$, $\sim 1/3$ are radio loud and $\sim 2/3$
radio quiet.

\begin{figure}
\plottwo{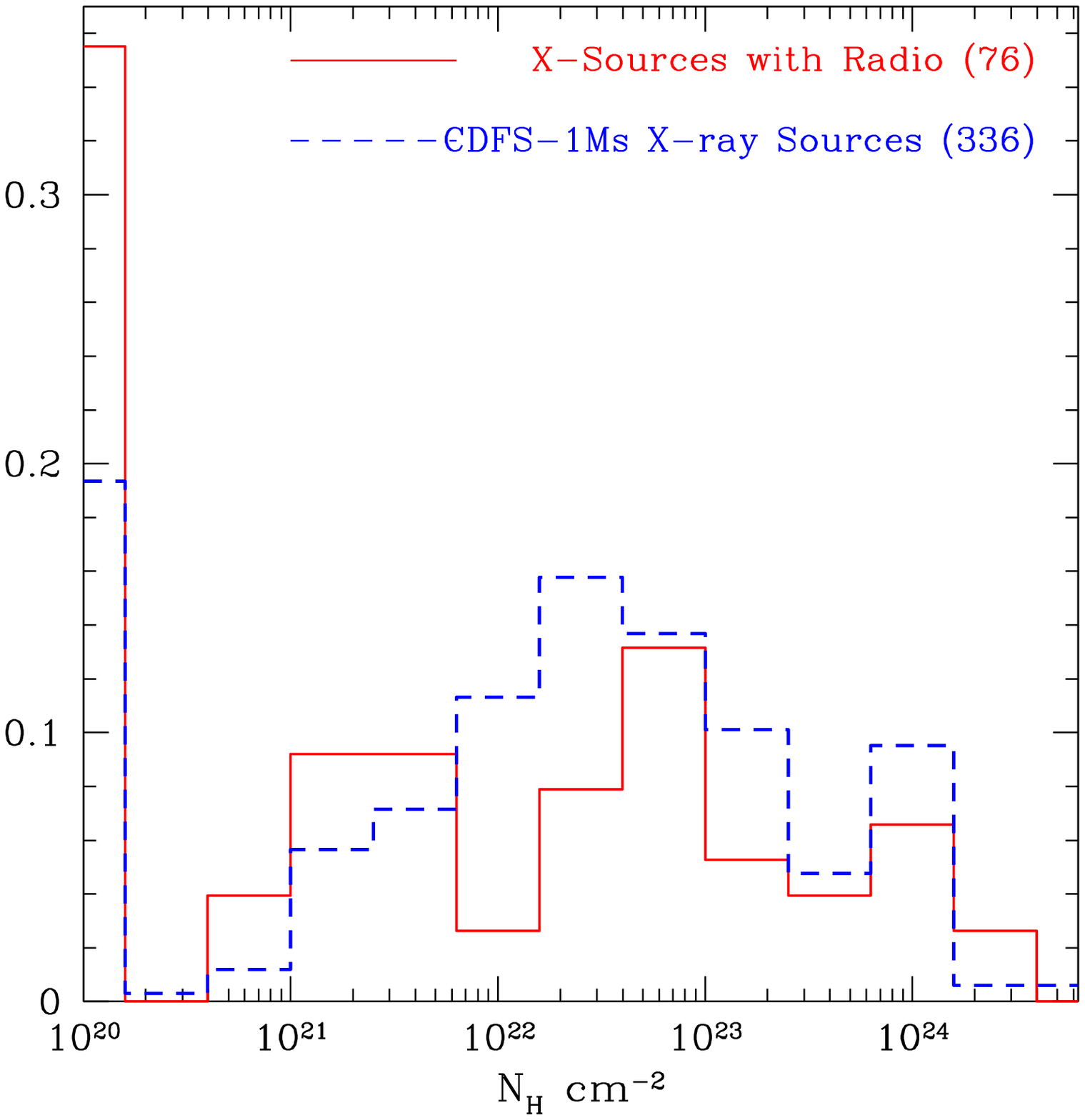}{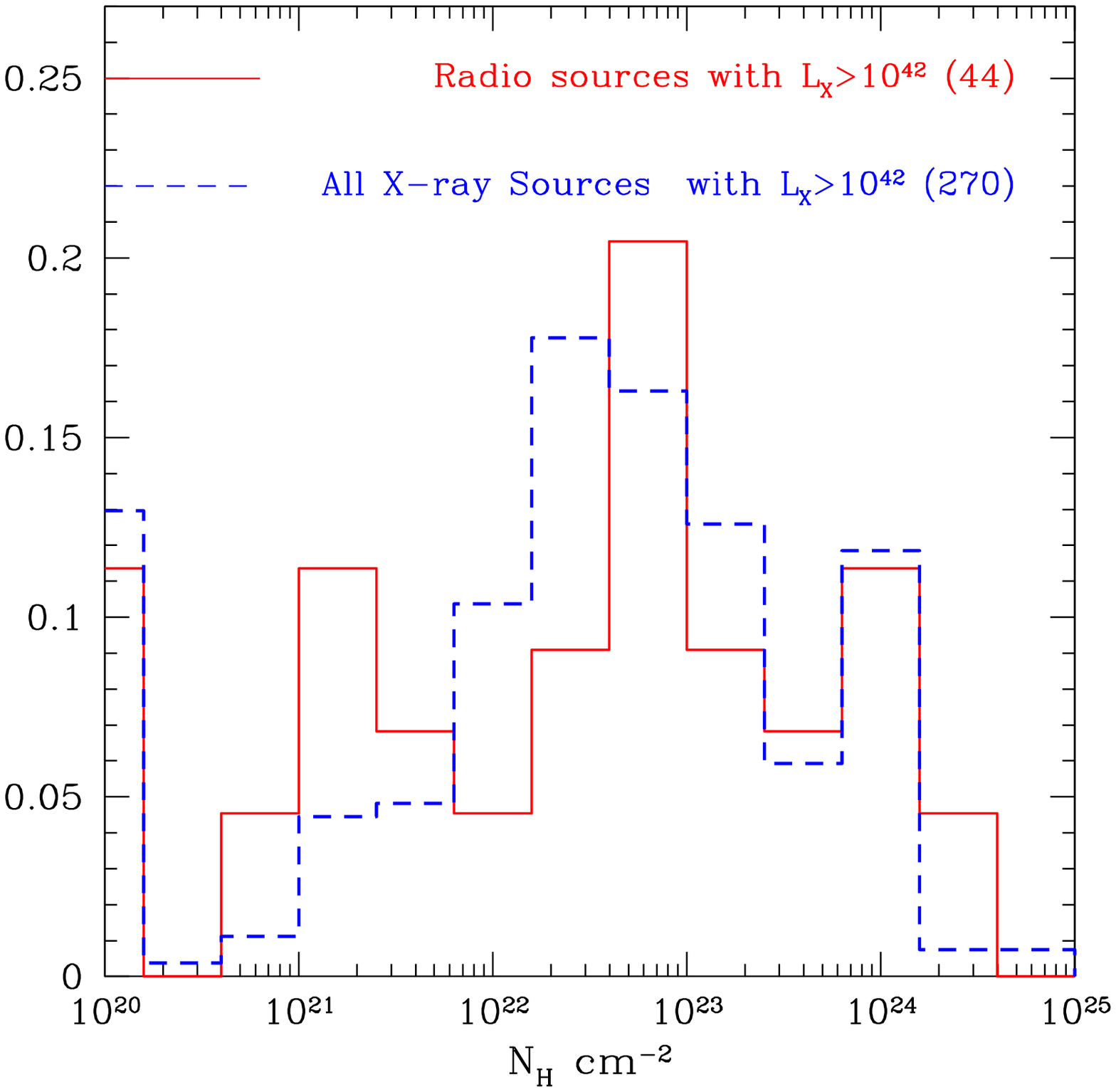}
\caption{{\sl Left}: Fractional distribution of measured intrinsic
  absorbing columns of equivalent $N_H$ for the X--ray sources with
  redshifts, with radio matches (continuous histogram).  The
  fractional distribution of absorbing columns of the entire X--ray
  sample is also shown (dashed histogram). The two distributions are
  inconsistent with each other.  {\sl Right}: same as in the left
  panel, but for sources with $L_X >10^{42}$ erg s$^{-1}$.  The two
  distributions are now consistent with each other. }
\label{nh}
\end{figure}

In Figure \ref{nh}, left panel, we plot the fractional distribution of
intrinsic absorbing columns of equivalent $N_H$ for the 76 radio
sources with X--ray spectral analysis and redshift information.  The
shape of the distribution for $N_H > 10^{21}$ cm$^{-2}$ is similar to
that of the entire X--ray sample (dashed line), but the two
distributions are inconsistent at more than 3 $\sigma$, due to the
significantly larger number of radio sources with low intrinsic
absorption.  This is reflected also in the larger fraction of sources
with $L_X<10^{42}$ erg s$^{-1}$ in the radio sample (35\%) compared
with the whole X--ray sample (20\%).  These X--ray sources are mostly
powered by starbursts, and therefore the X--ray emission does not show
the intrinsic absorption found in sources with nuclear emission.

We divided sources into two subsamples according to the X--ray
luminosity, and we find that radio sources with $L_{2-10}> 10^{42}$
erg s$^{-1}$ have a distribution of $N_H$ consistent with that of
X--ray selected sources in the same luminosity range (see Figure
\ref{nh}, right panel).  Again, this shows that the subsample of radio
sources with X--ray luminosity $L_{2-10}> 10^{42}$ erg s$^{-1}$ is
representative of the X--ray selected AGN population.  This does not
show any significant difference in the intrinsic absorption properties
of X-ray sources with and without radio counterpart.  This allows one
to discard a simple model in which the radio emission is associated to
starbursts which in turn would absorb the X--ray emission from the
AGN, as discussed by Rovilos et al. (2007).  As a general remark, we
recall that the gap in the distribution at $N_H < 10^{21}$ cm$^{-2}$
is not due to difficulties in measuring low values of $N_H$,
especially at high redshifts.  Indeed, as shown in Figure 9 of Tozzi
et al. (2006), the resampling of $N_H$ values according to the
statistical error, decreases only by 20\% the number of sources with
$N_H < 10^{21}$ cm$^{-2}$.  Here, for simplicity, we do not correct
for this effect, nor for incompleteness (see, again, Tozzi et
al. 2006) since we mostly focus on the comparison between radio
sources with X--ray counterpart and the parent X--ray population.

\begin{figure}
\plotone{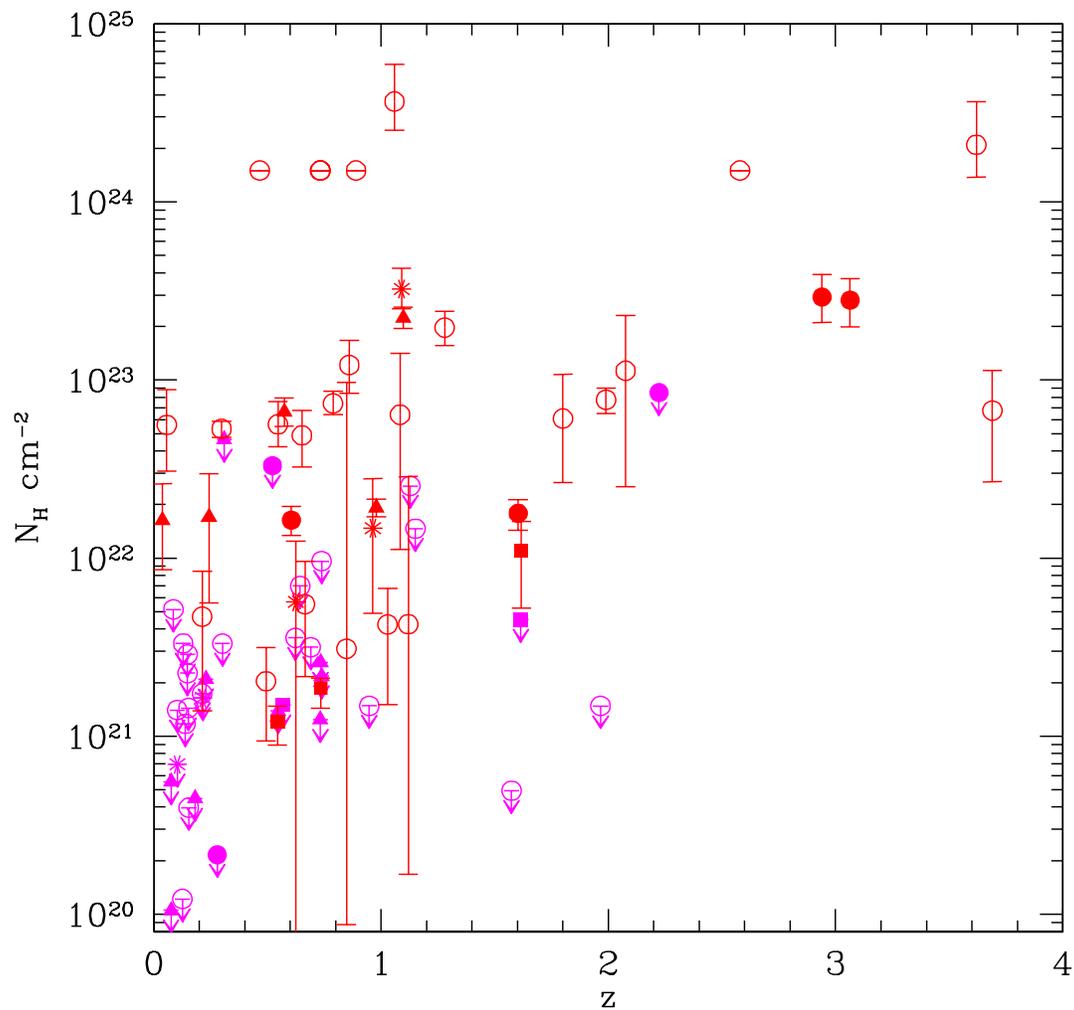}
\caption{Intrinsic absorption versus redshift for the X--ray radio
  matches.  Upper limits (1 $\sigma$) are used for measures consistent
  with $N_H = 0$ within 1 $\sigma$.  Compton thick candidates are
  plotted at $N_H = 1.5 \times 10^{24}$ cm$^2$ as lower limits to the
  actual value.  Error bars correspond to 1 $\sigma$. Different
  symbols as in Figure \ref{optype} (empty circles for sources
  without optical spectral classification).  }
\label{nhz}
\end{figure}

In Figure \ref{nhz} we plot the intrinsic absorption versus redshift.
The apparent increase of $N_H$ with redshift is due to the difficulty
of measuring $N_H$ at high redshift as discussed in Tozzi et
al. (2006).  In this Figure we see clearly that the large number of
$N_H$ upper limits, causing the difference in figure \ref{nh} (left
panel), are mostly at low redshifts.  The two effects are clearly the
same, and are due to the X--ray flux limit, which introduces a sharp
cutoff around $z\sim 1$ for sources with $L_X \leq 10^{42}$ erg
s$^{-1}$, where all the star forming galaxies, showing no intrinsic
absorption, are found.

\begin{figure}
\plotone{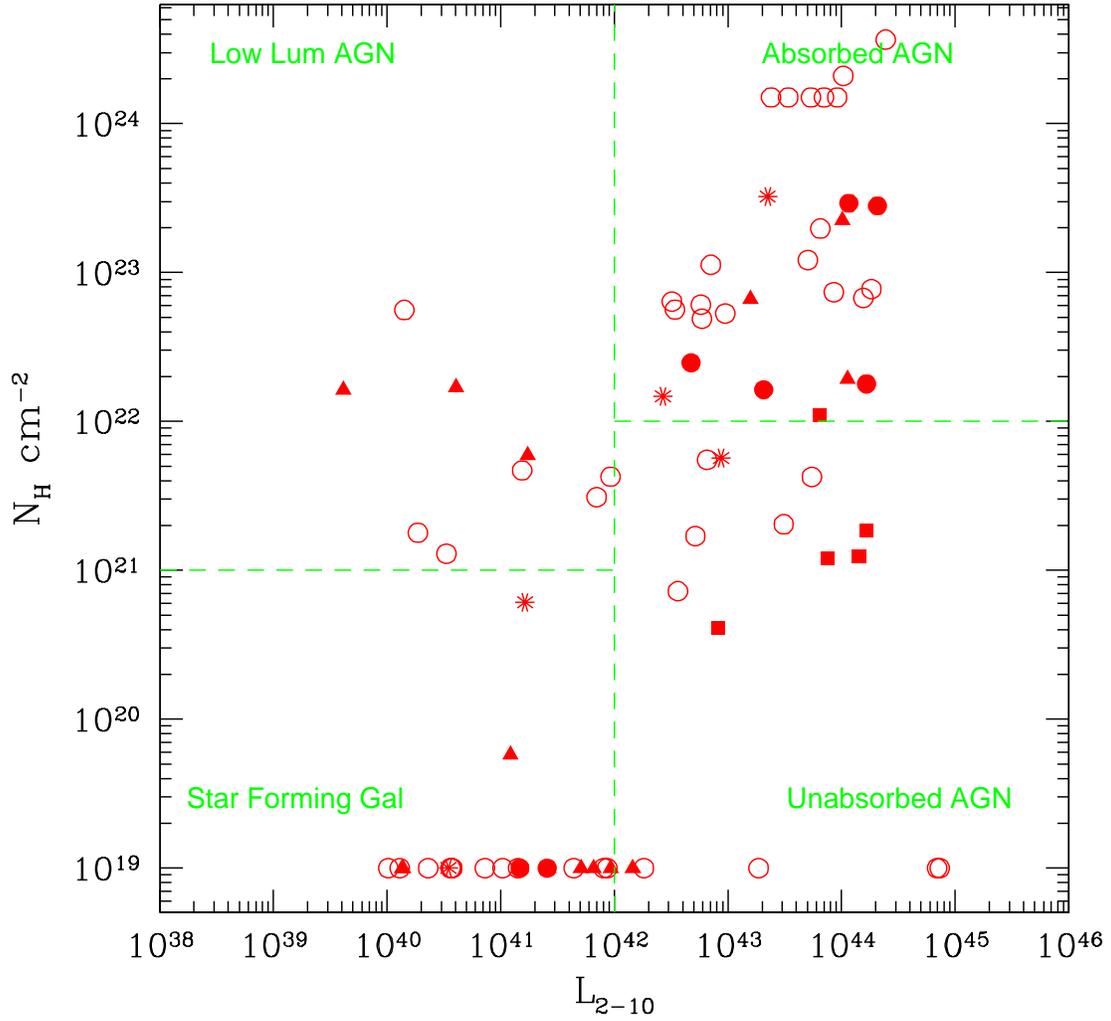}
\caption{Intrinsic absorbing columns of equivalent $N_H$ versus
  intrinsic hard luminosities for the X--ray radio matches.  Different
  symbols as in Figure \ref{optype} (empty circles for sources
  without optical spectral classification).  Dashed lines provide a
  simple X--ray classification as described in the text.}
\label{xclass}
\end{figure}

To summarize the properties of X--ray detected radio sources, we show
in Figure \ref{xclass} a simple but efficient classification based on
X--ray properties only.  We assume $L_{2-10} = 10^{42}$ erg s$^{-1}$
as the threshold luminosity separating star forming galaxies and AGN,
and $N_H = 10^{22}$ cm$^{-2}$ as the conventional threshold intrinsic
absorption for unabsorbed and absorbed AGN.  We find 23 star forming
galaxies, 15 unabsorbed AGN and 29 absorbed AGN. In the star forming
regime, the presence of absorption, not necessarily with high $N_H$,
is the signature of nuclear emission.  Therefore, if we adopt $N_H =
10^{21}$ cm$^{-2}$ as a conservative threshold, we can also
tentatively identify 9 low luminosity AGN at $L_{2-10} < 10^{42}$ erg
s$^{-1}$.  The optical classification for 40\% of the sources show
that several AGN are missed by optical spectroscopy, while only two
HEX are included in the star forming galaxies sector.

\begin{figure}
\plotone{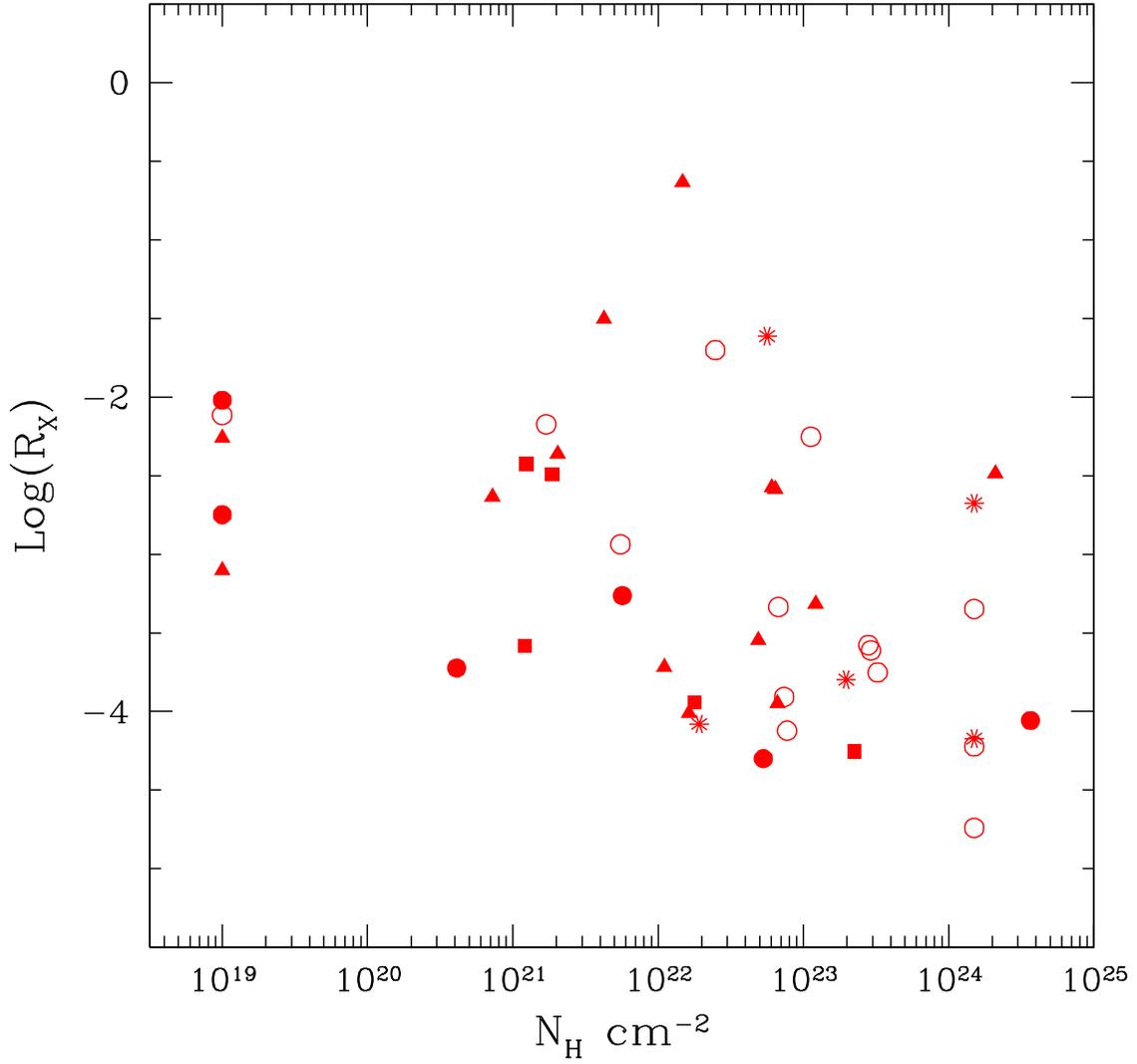}
\caption{Radio loudness $R_X$ versus the intrinsic absorption for
  sources with $L_{2-10} > 10^{42}$ erg s$^{-1}$.  Same symbols as in
  Figure \ref{optype} (empty circles for sources without optical
  spectral classification). }
\label{rx_nh}
\end{figure}

Broadly speaking, we find that X--ray emission of 1/3 of the X--ray
detected radio sources is consistent with being associated to star
formation in the host galaxy, while the remaining 2/3 are AGN.  We
also find a weak correlation between the radio loudness and the
intrinsic absorption among the sources with $L_{2-10} > 10^{42}$ erg
s$^{-1}$.  As shown in Figure \ref{rx_nh}, there is a large scatter,
but the radio loudness is significantly higher at lower $N_H$.  The
Spearman rank correlation coefficient is -0.2, with a significance
weaker than 2 $\sigma$.  This is a hint that radio emission is
decreasing with increasing absorption among X-ray detected AGN.

\begin{figure}
\resizebox{\hsize}{!}{\includegraphics{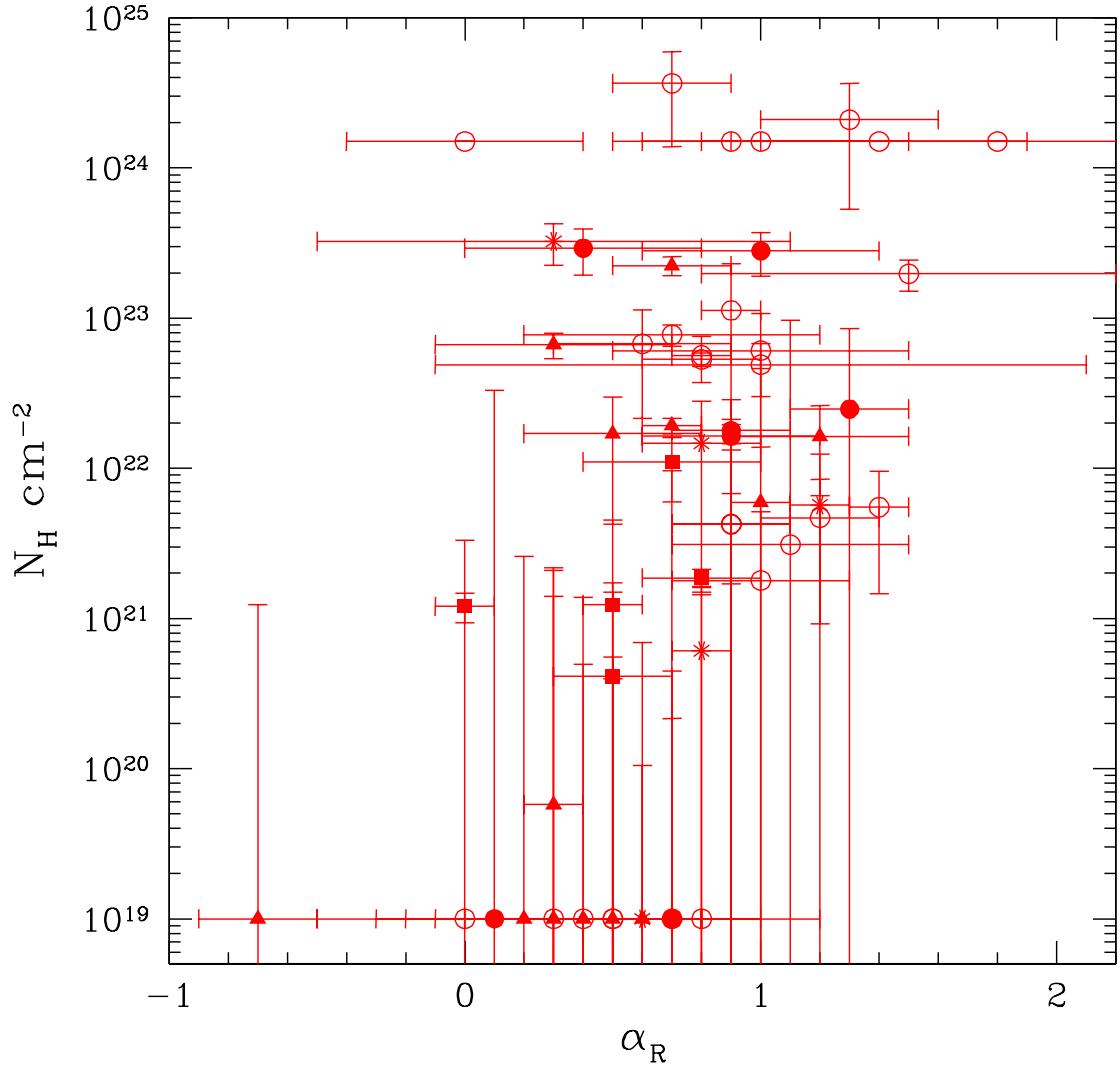}}
\caption{Radio spectral slope $\alpha_R$ versus the intrinsic
  absorption $N_H$.  Same symbols as in Figure \ref{optype} (empty
  circles for sources without optical spectral classification).}
\label{alpha_nh}
\end{figure}

Finally we find a correlation between the radio spectral index
$\alpha_R$, computed between 20 and 6 cm, and the intrinsic absorption
measured in X--ray.  The correlation is significant at the 3 $\sigma$
level in a Spearman rank correlation test for the 59 sources which
have both measured 6 cm flux densities and X--ray spectral analysis.
The relation is shown in Figure \ref{alpha_nh} for only 59 sources
which have both measured 6 cm fluxes and X--ray spectral analysis.  On
the other hand, we do not find a correlation between the radio
spectral index and hard X--ray luminosity, nor between radio spectral
index and radio-X--ray loudness.  The apparent trend of having flatter
radio spectra at lower intrinsic absorption may be due to a component
of thermal radio emission dominating at low $N_H$ due to star
formation processes, while AGNs, with significant intrinsic
absorption, show steeper radio spectra typical of non thermal
transparent synchrotron emission (see Richards 2000).  Indeed, the
correlation we found in our sample is partially due to the presence of
star forming galaxies, since it becomes significant only at the 90\%
level when only sources with $L_{2-10}> 10^{42}$ erg s$^{-1}$ are
included.  Finally, we notice that the average X--ray spectral slope
of the X--ray detected radio sources is $\Gamma = 1.8 \pm 0.1$, in
agreement with that of the X--ray sample.

\section{Radio sources without X--ray counterparts}

We have 174 radio sources in the FOV of the CDFS 1Ms exposure or in
the complementary area of the E-CDFS without an X--ray counterpart (we
note that the sources with RID = 1, 21 and 266 are outside the E-CDFS
field).  To retrieve X--ray information about these sources, we
performed aperture photometry on the X--ray images at the radio
positions.  We use X--ray images obtained by masking the cataloged
X--ray sources, replacing the removed regions with a Poissonian
background based on the measured value of the local background.  In
this way we avoid including the emission from any detected X--ray
source.  We performed the photometry separately in the soft (0.5--2
keV) and hard (2--7 keV) X--ray bands.  The results are given in Table
\ref{table4}.  The net counts and the signal--to--noise are computed
in the extraction radius which is dependent on the off--axis angle of
the X--ray images, as described in Giacconi et al. (2001).

As we see from Table \ref{table4}, there are 26 sources whose S/N
ratio from aperture photometry in one of the two bands, is higher than
the S/N limit for X--ray detect sources.  However, these sources
should not be considered X--ray detections since, on the basis of the
X--ray detection algorithm, they have a very low probability of being
real sources, and therefore were not included in the Giacconi et
al. (2002) or the Lehmer et al. (2005) catalogs.  Eventually, we have
included their contribution in the stacking analysis of all the X--ray
undetected radio sources.  For all the other sources, we quote the 3
$\sigma$ upper limits both in counts and fluxes.

\begin{figure}
\plottwo{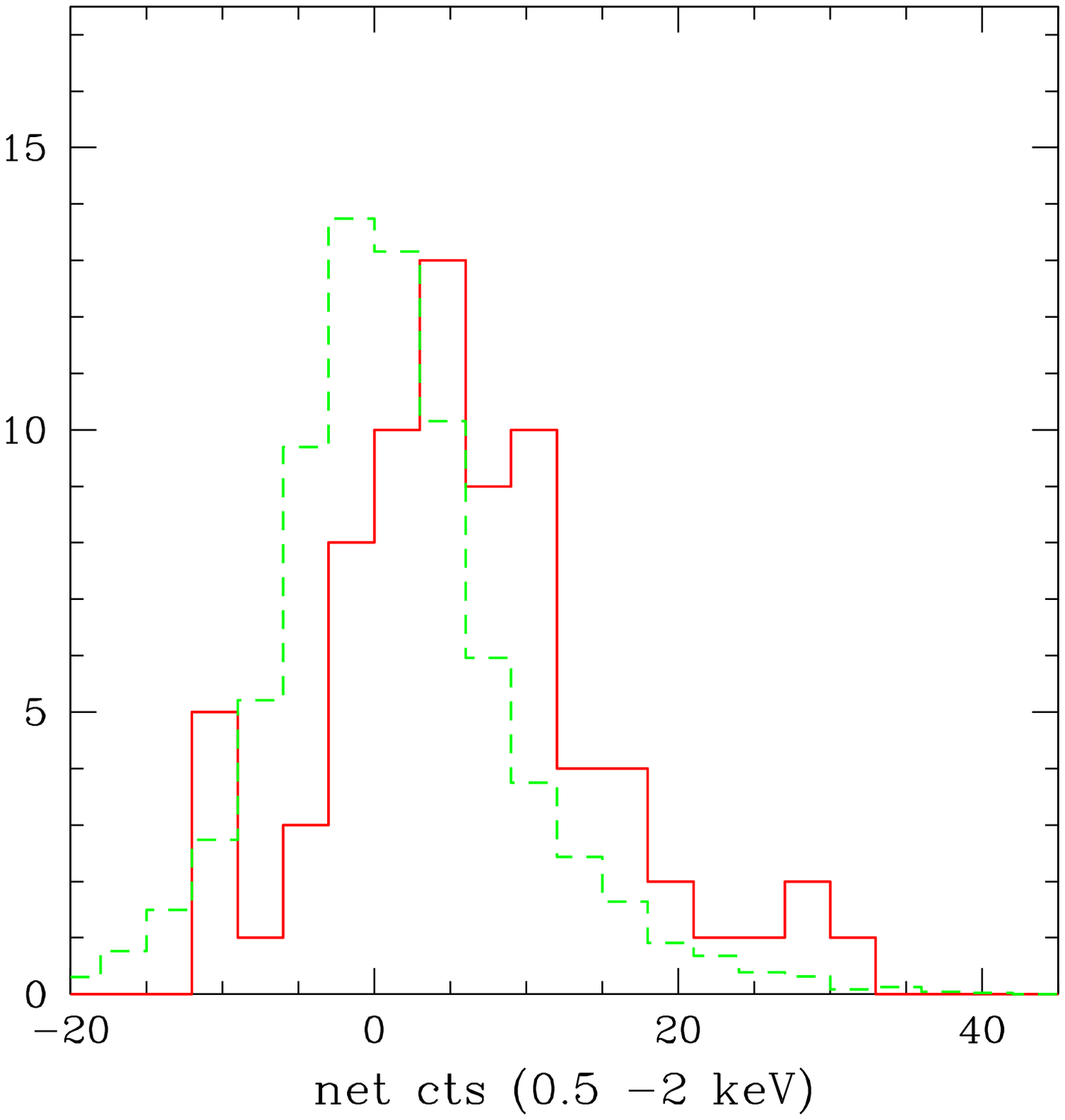}{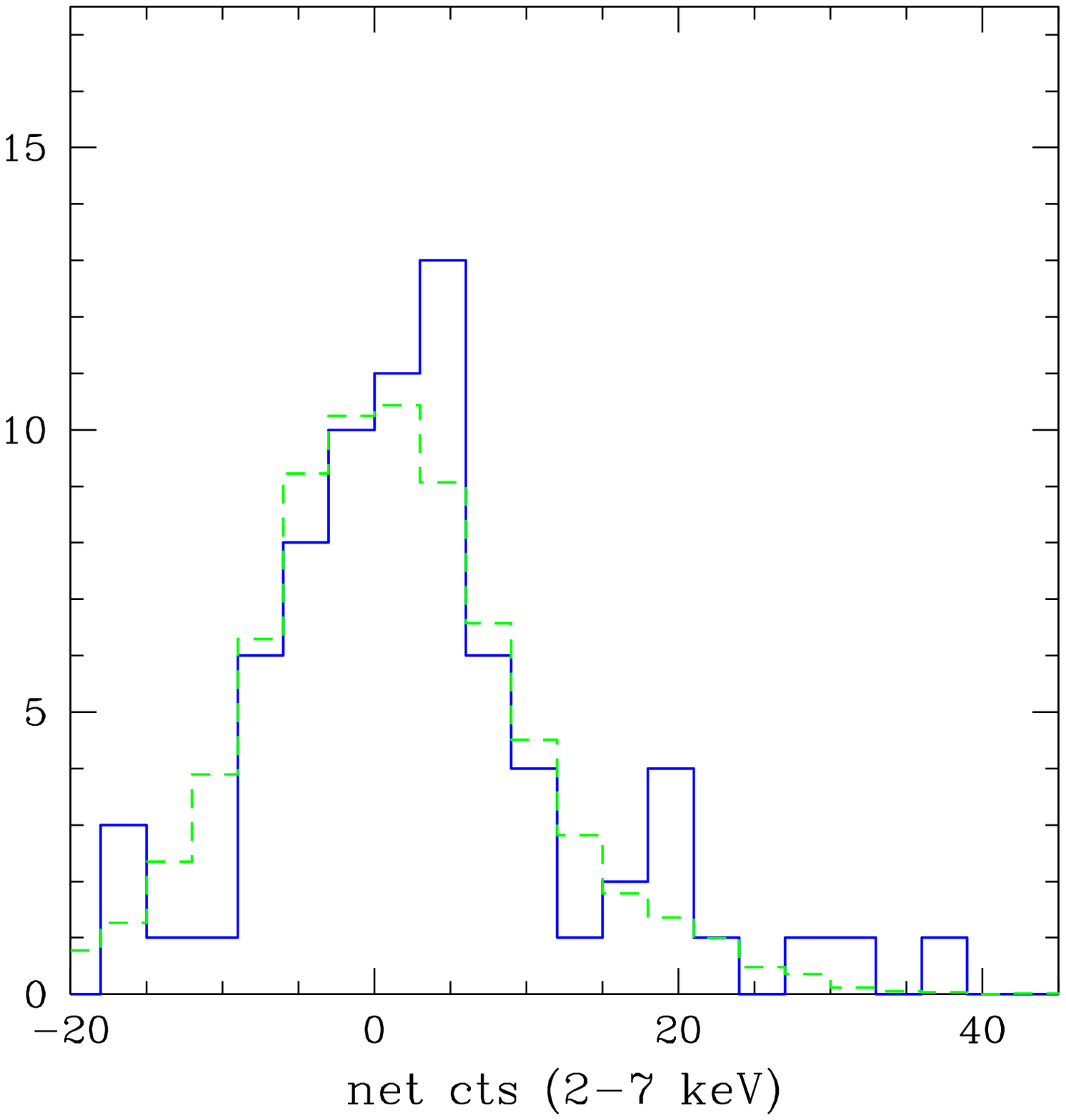}
\caption{{\sl Left}: Histogram distribution of the net counts in the
  1Ms exposure for radio sources without X--ray counterparts in the
  soft band (solid line) compared with random photometry (dashed
  line).  {\sl Right}: the same in the hard band.  }
\label{netcounts_1M}
\end{figure}

\begin{figure}
\plottwo{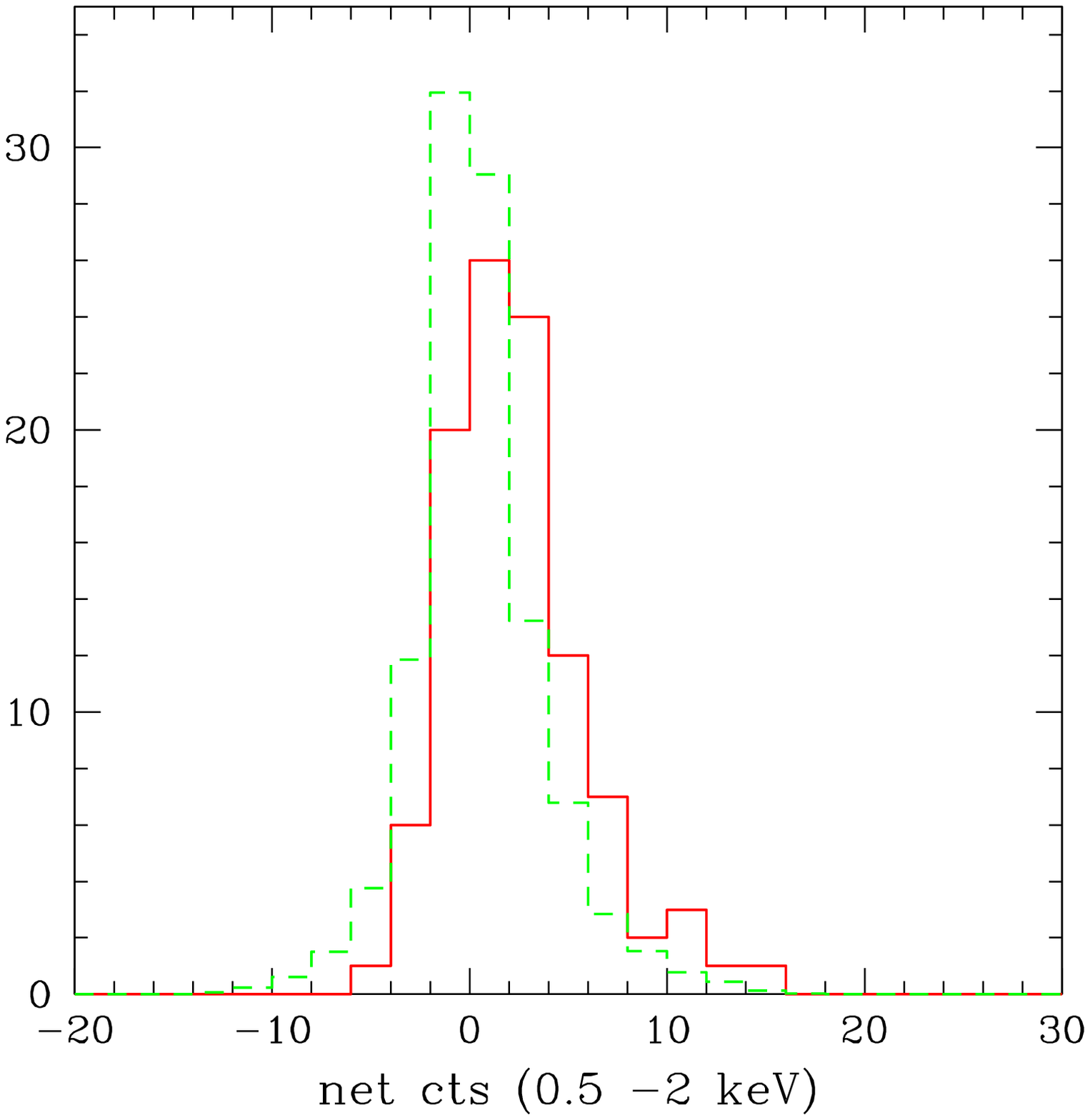}{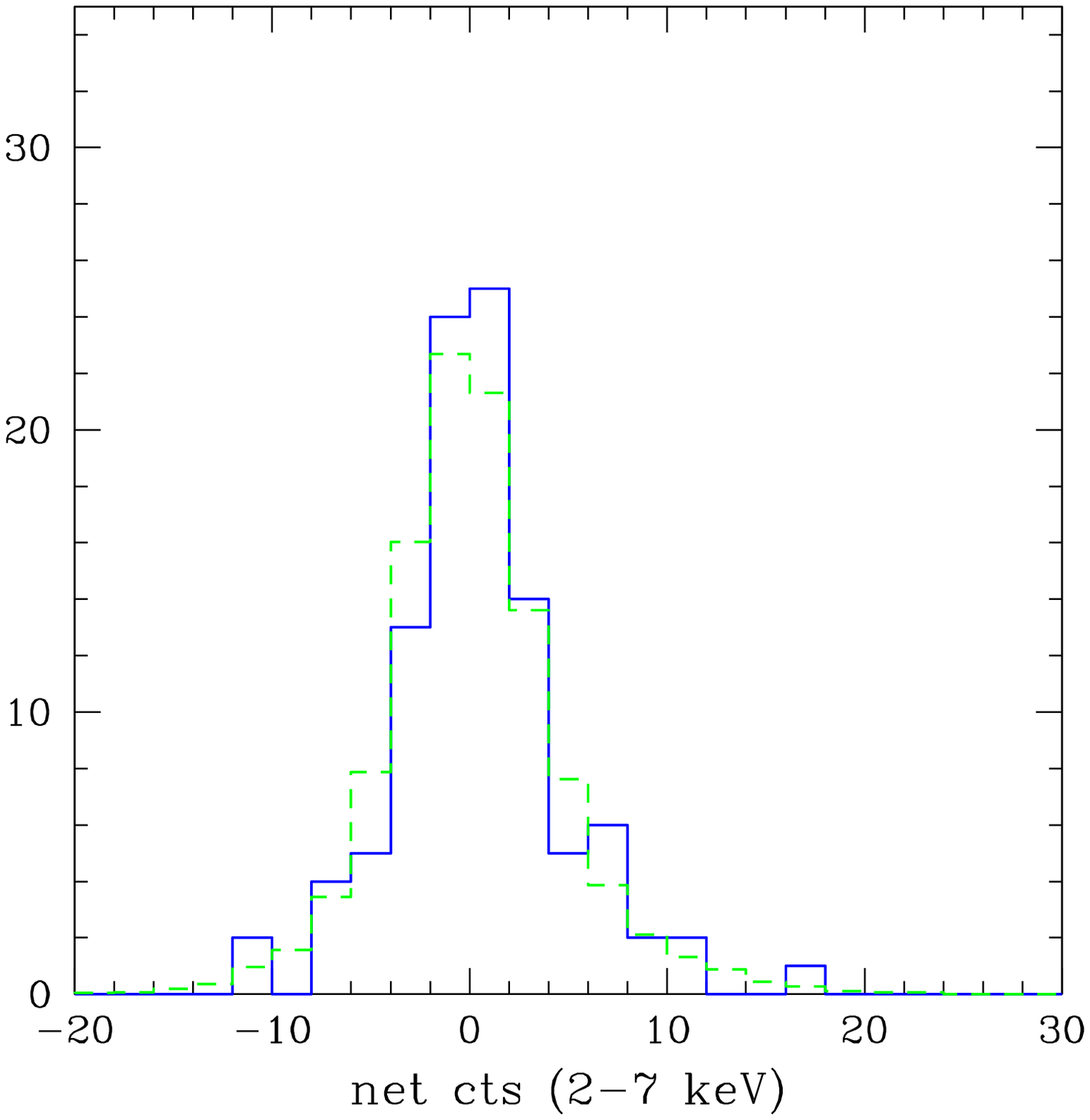}
\caption{{\sl Left}: Histogram distribution of the net counts in the
  complementary E-CDFS exposure for radio sources without X--ray
  counterparts in the soft band (solid line) compared with random
  photometry (dashed line).  {\sl Right}: the same in the hard band.
}
\label{netcounts_EXT}
\end{figure}

In general, the histogram distributions of the net counts, shown
separately in the 1Ms exposure (Figure \ref{netcounts_1M}) and in the
complementary area of the E-CDFS (Figure \ref{netcounts_EXT}), show a
clear excess with respect to a distribution of photometry based on
random positions (dashed line) in the soft band (left panel), and a
marginal excess in the hard band (right panel).  The probability that
the measured and random net counts distributions are different is more
than 5 sigma
in the soft band, while it is marginal in the hard band for sources in
the 1Ms CDFS (80\%) and in the E-CDFS (60\%).  In any case, there are
no sources which dominate the X--ray photometry of the radio sources
with no cataloged X--ray sources, allowing us to perform a meaningful
stacked analysis to obtain their average properties.

\begin{figure}
\plottwo{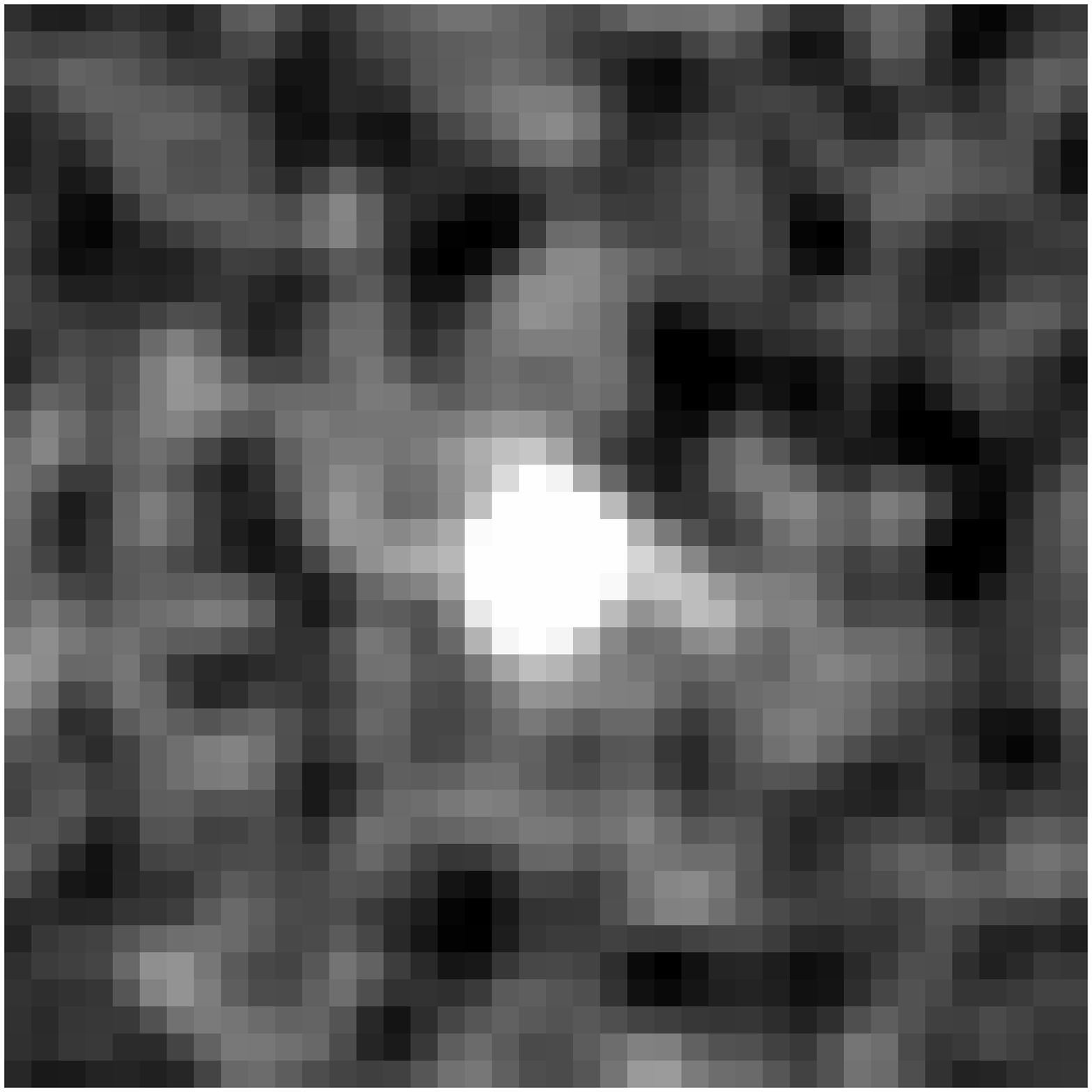}{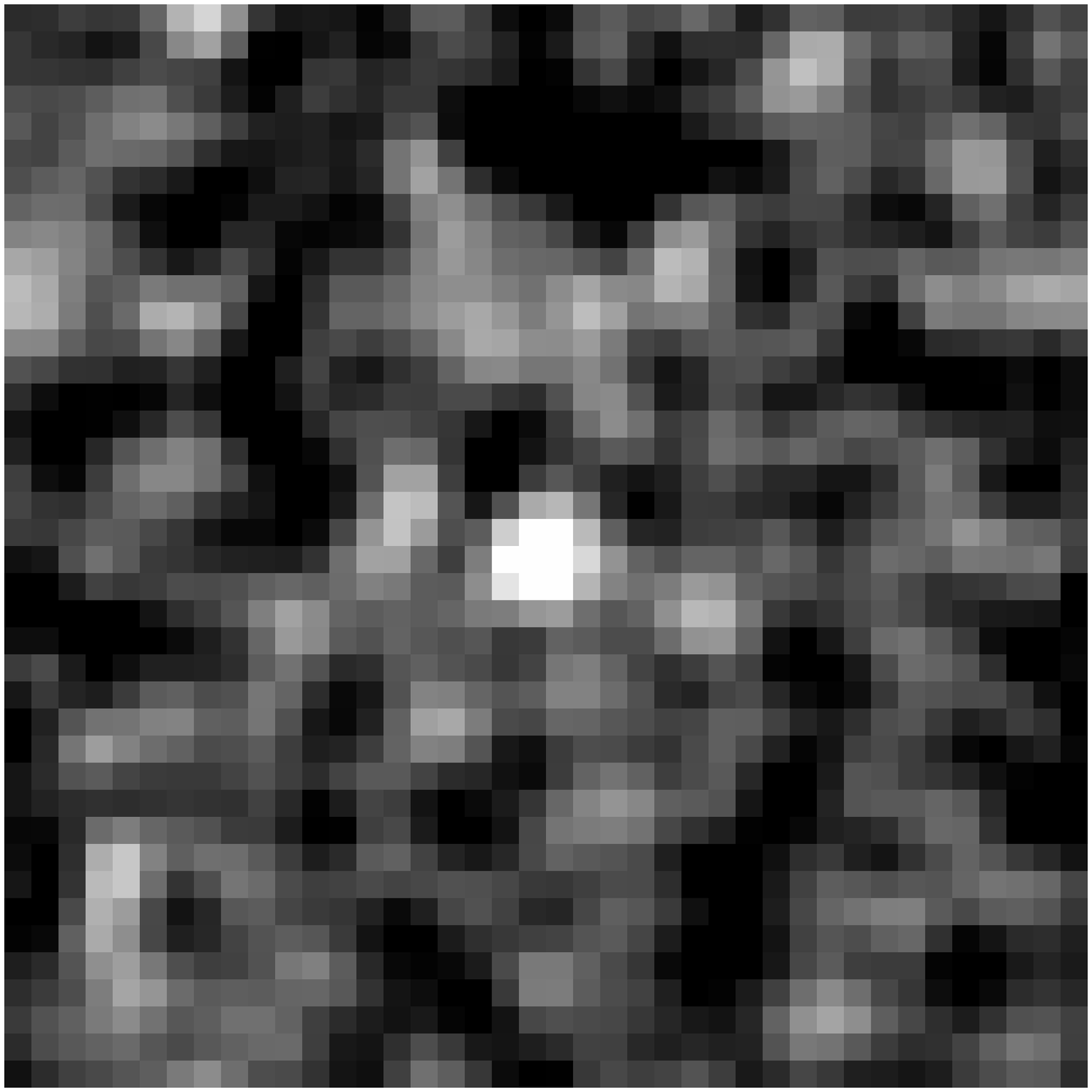}
\caption{Stacked images of the 74 radio sources without X--ray
  counterparts in the 1Ms field in the soft (left) and hard (right)
  bands. }
\label{stacked_1Ms}
\end{figure}

A visual impression of the total X--ray emission from the radio
sources without counterparts in the X--ray catalog can be obtained
simply by stacking the X--ray image at the position of the radio
sources.  The stacked images in the soft and hard bands are shown in
Figure \ref{stacked_1Ms}.  Overall, the 74 radio sources in the 1Ms
field are detected with $ 460 \pm 75$ and $300 \pm 90$ net counts in
the soft and hard bands respectively.  We performed a Monte Carlo
simulation to assess the significance of the detection of the stacked
image.  The detection in the soft band is at more than 99.9\%
confidence level, while in the hard band it is at 99\%.  Energy fluxes
are computed as in Rosati et al. (2002), using conversion factors from
the measured net count rate to the energy flux, assuming an average
spectral slope of $\Gamma = 1.4$, equal to $5.07\times 10^{-12} $ and
$ 2.97\times 10^{-11}$ erg s$^{-1}$ cm$^{-2}$ (cnts s$^{-1}$)$^{-1}$
in the soft and hard bands respectively.
After correcting for the effective average X--ray exposure time, the
photometry of the stacked images corresponds to a flux of $(4.4 \pm
0.7) \times 10^{-17}$ erg s$^{-1}$ cm$^{-2}$ and $(1.7 \pm 0.3) \times
10^{-16}$ erg s$^{-1}$ cm$^{-2}$ per source in the soft and hard bands
respectively.

\begin{figure}
\plottwo{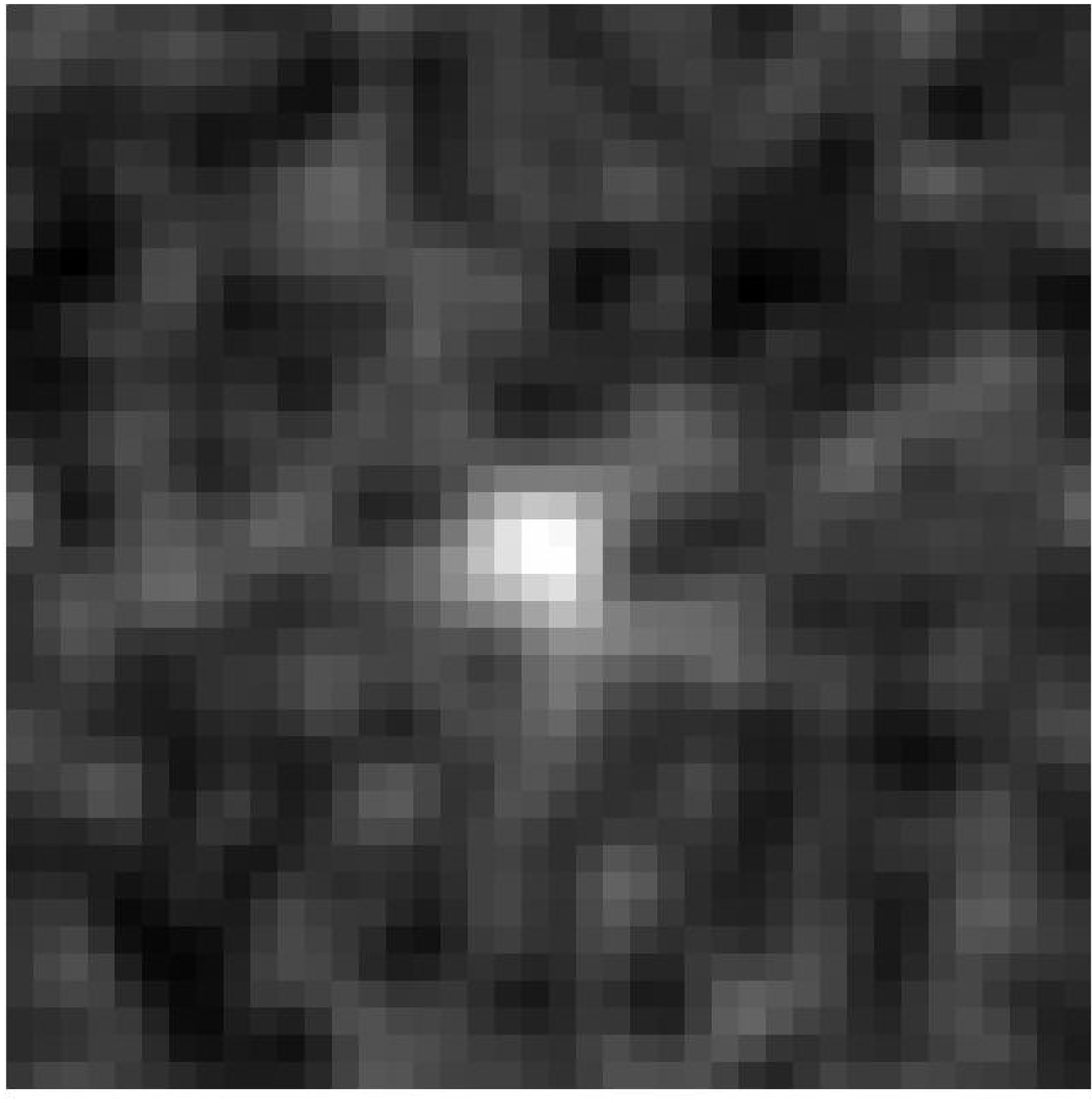}{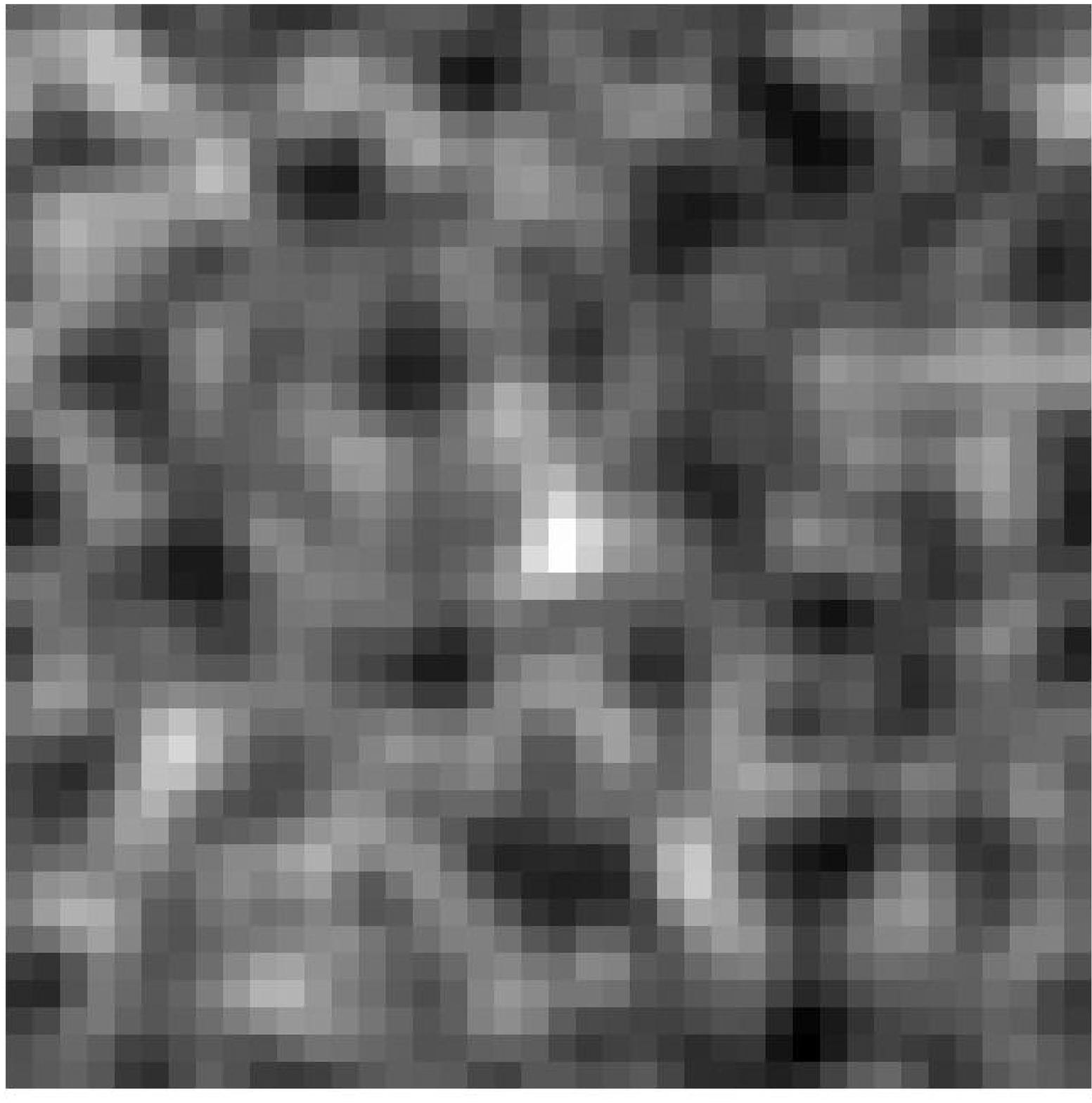}
\caption{Stacked images of the 103 radio sources without X--ray
  counterparts in the E-CDFS field in the soft (left) and hard (right)
  bands. }
\label{stacked_EXT}
\end{figure}

With the same analysis, the 100 radio sources in the E-CDFS field
yield $ 260 \pm 30$ and $90 \pm 60$ net counts in the soft and hard
bands respectively.  The stacked images in the soft and hard bands are
shown in Figure \ref{stacked_EXT}.  The detection in the soft band is
highly significant, while in the hard band it is only marginal.  After
correcting for the effective average exposure, and adopting the
appropriate conversion factors for the E-CDFS fields, the average flux
per source is $(5.4 \pm 0.6) \times 10^{-17}$ erg s$^{-1}$ cm$^{-2}$
and $(1.7 \pm 1.1) \times 10^{-16}$ erg s$^{-1}$ cm$^{-2}$ in the soft
and hard bands respectively.  These values are consistent with those
found in the CDFS, supporting the accuracy of the stacking procedure.


To investigate further the nature of the weak X--ray emission, we
evaluate the average hardness ratio defined as $HR = (H-S)/(H+S)$,
where $H$ and $S$ are the net counts in the hard (2--7 keV)
and soft (0.5--2 keV) bands respectively, corrected for vignetting.
If we stack the net counts  of sources in four bins of radio
flux density, we find a roughly constant value of $HR \sim -0.5 \pm
0.1$, indicating that the statistics are not able to indicate a
significant change in the average X--ray spectral properties of the
radio sources as a function of their radio flux density.

\begin{figure}
\plotone{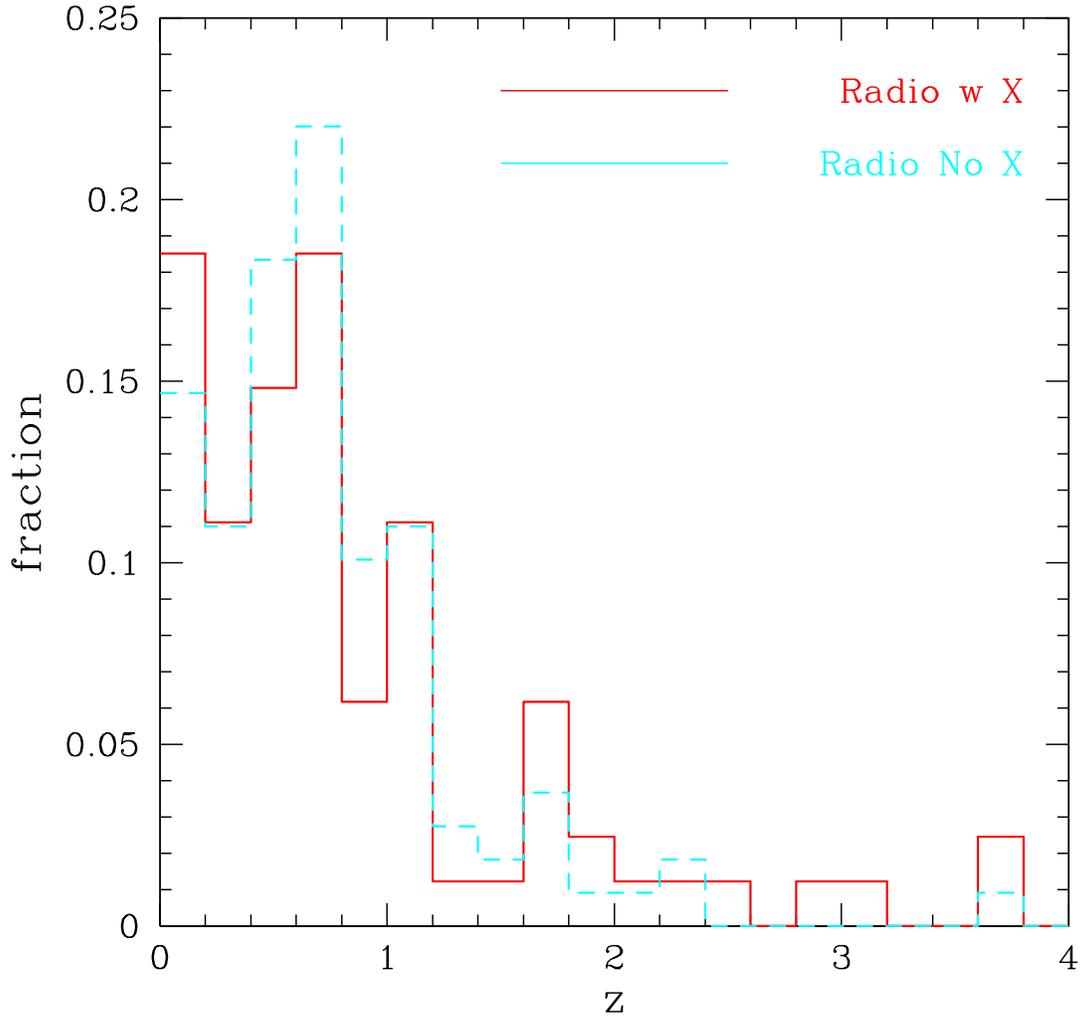}
\caption{Redshift distribution of the 76 radio sources with X--ray
  counterparts and redshift information (solid line) compared with the
  redshift distribution of the 110 radio sources without X--ray
  counterparts and measured redshift (dashed line).  }
\label{z_onlyradio}
\end{figure}
\begin{figure}
\plotone{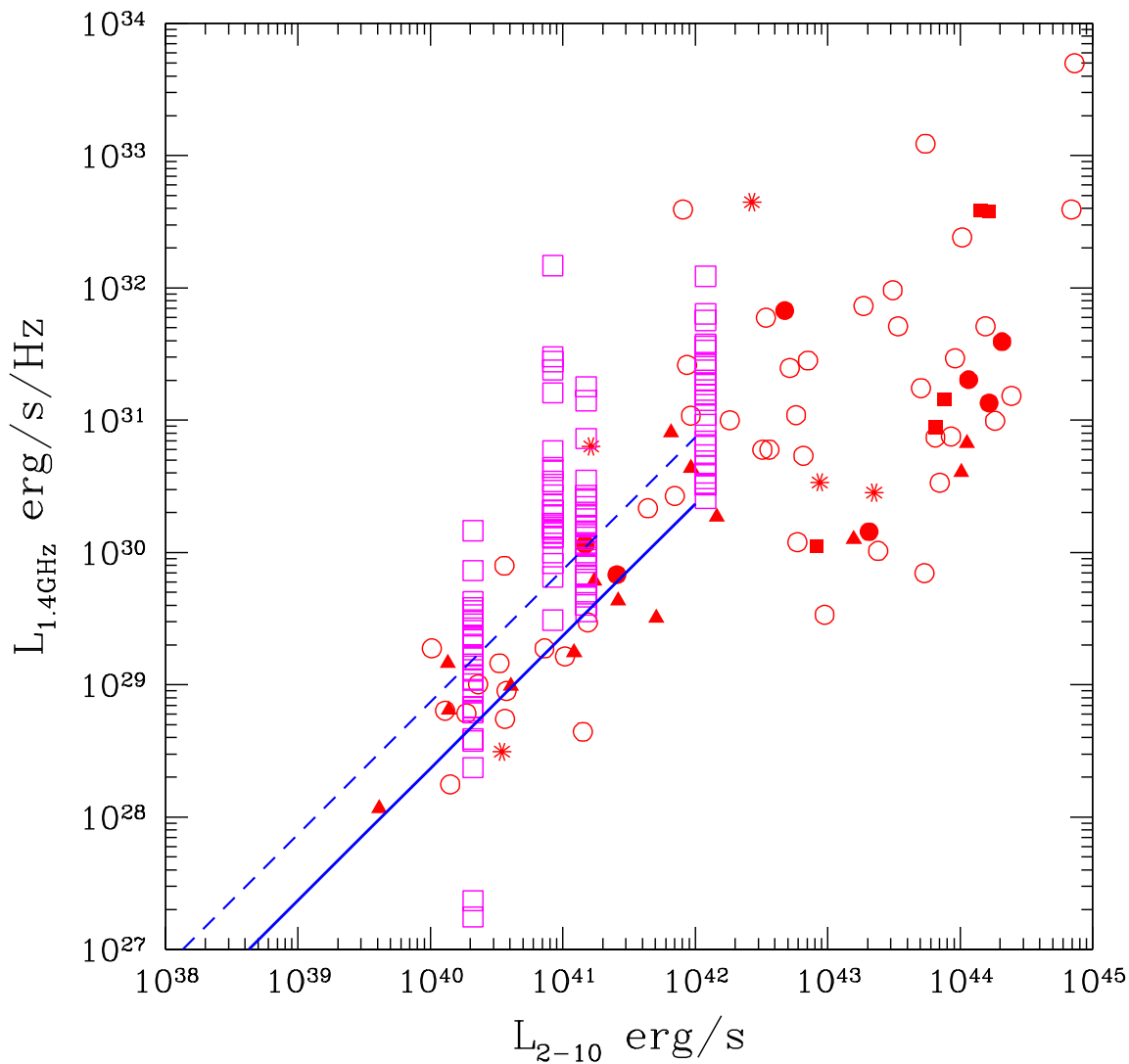}
\caption{Radio luminosity plotted against the X--ray luminosity in the
  hard bands (blue triangles) as in Figure \ref{lum}, with the radio
  luminosities and average X--ray luminosities of radio--only sources
  shown as empty square, computed for sources in four redshift bins:
  $0.0<z<0.4$, $0.4 < z< 0.7$, $0.7 < z < 1$, $ 1 < z < 2.3$.  Lines
  as in Figure \ref{lum}.  Same symbols as in Figure \ref{optype}
  (empty circles for sources without optical spectral
  classification). }
\label{or_lum}
\end{figure}

We have redshifts for 64\% (110) of our radio sources with no X--ray
detection.  The redshift distributions of the sources with and without
X--ray counterparts are consistent with each other, as shown in Figure
\ref{z_onlyradio}.  If we split the sample in four redshift bins with
about 27 sources each, we can measure the average X--ray luminosities
for radio sources, using the average fluxes, and assuming a power law
spectrum with $\Gamma = 1.8$.  The four redshift bins are:
$0.0<z<0.4$, $0.4 < z< 0.7$, $0.7 < z < 1$, $ 1 < z < 2.3$.  The
results, listed in Table \ref{table5}, show that the X--ray luminosity
for these sources lie in the range of star forming galaxies.  Only in
the higher redshift bin ($\langle z \rangle \sim 1.4$) is the average
hard X--ray luminosity at the high end of the typical starburst
galaxies.  However, to evaluate the contribution from a population of
low luminosity AGN, a multiwavelength approach, as discussed in Paper
IV, is needed.  If we plot these sources in the $L_R$--$L_X$ plane, we
find that, on average, they are consistent with the relation expected
for star forming galaxies (see Figure \ref{or_lum}).  This is
consistent with the censored analysis presented in Paper IV,
confirming the results by Lehmer et al. (2008) supporting the
$L_X$--$L_R$ correlation holding at high redshift; this is at variance
with the claim of Barger et al. (2007).


\section{Conclusions}

We present detailed X-ray spectral properties of 76 VLA sources with
X-ray counterparts in the CDFS (Giacconi et al. 2002) and the E-CDFS
(Lehmer et al.  2005).  We also present the average X--ray properties
of the radio sources without X--ray counterparts in the 1Ms CDFS
exposure and in the E-CDFS field.  Our main results are summarized as
follows.

\begin{itemize}

\item One third of the radio sources are detected in the X--ray bands.
  Among them, $\sim 1/3$ of the radio sources are consistent with
  being star forming galaxies, while the remaining 2/3 are AGN, by
  assuming $L_X = 10^{42}$ erg s$^{-1}$ as the threshold between star
  forming galaxies and AGN.

\item In the AGN luminosity range, $L_{2-10} > 10^{42}$ erg s$^{-1}$,
  $\sim 1/3$ of the sources are radio loud and $\sim 2/3$ radio quiet,
  where radio loud is defined as $log(R_X)>-2.9$ (with $R_X \equiv \nu
  L_R(5 GHz) /L_{2-10}$).

\item The intrinsic absorption in the X--ray band of the radio sources
  is shifted to lower values with respect to the X--ray selected
  sample, showing that radio selection tends to find a larger number
  of star forming galaxies; when selecting source with $L_{2-10}>
  10^{42}$ erg s$^{-1}$ the distribution is similar to that of the
  X--ray sample.

\item We find a weak anticorrelation of radio loudness as a function
  of intrinsic absorption, adding support to the finding that radio
  emission is not efficient in selecting more absorbed AGN.

\item The stacked X--ray images of 174 radio sources without cataloged
  X--ray counterparts shows a clear detection in the soft band and a
  marginal detection in the hard band.

\item The average X--ray luminosities of radio sources without
  cataloged X--ray counterpart is consistent with being powered by
  star formation.

\end{itemize}

Deeper X--ray and radio data in the CDFS will allow us to extend this
analysis toward lower levels, and to obtain additional contraints on
the role of star forming galaxies as opposed to AGN in the sub mJy
radio population.

\acknowledgements

P. Tozzi acknowledges support under the ESO visitor program in
Garching during the completion of this work.  We thank Massimo Persic
and Piero Ranalli for discussion on the X--ray/Radio luminosity
correlation, and Isabella Prandoni for valuable comments.  We
acknowledge financial contribution from contract ASI--INAF I/023/05/0
and from the PD51 INFN grant.  The VLA is a facility of the National
Science Foundation operated by NRAO under a cooperative agreement with
Associated Universities Inc.

\clearpage


\clearpage

\begin{deluxetable}{rrrrrrrrr}
\tablewidth{0pc}
\tablecaption{Optical and X--ray properties of Radio sources with X--ray
  matches in the 1Ms exposure.  $Q$ refers to the optical spectrum
  quality (see Zheng et al. 2004).  The optical type has the following
  meaning: 1 = Broad Line AGN (BLAGN); 2= High Excitation line (HEX);
  3 = Low Excitation Lines (LEX); 4 = Absorption spectrum galaxy
  (ABS).  Optype=0 is when only photometric redshifts are available or
  no optical classification is possible.  Sources XID=29 and XID=51
  are fitted with a soft component, and sources XID=502, 608, and 913
  are fitted with a reflection model (pexrav) as in Tozzi et
  al. (2006). \label{table1}
}
\tablehead{
R ID & X ID & z & Q & Optype & $\Gamma$ & $N_H$ $10^{22}$ cm$^{-2}$ &  $L_{0.5-2}$ erg s$^{-1}$ & $L_{2-10}$ erg s$^{-1}$ \\}
\startdata
54	&	112	& 2.9400 & 2 &  2 & $ 1.8 $ &  $ 29.2_{-8.2}^{+9.9} $ & $  7.50 \times 10^{43} $  & $ 1.16 \times 10^{44} $  \\ 
66	&	74	& 0.6650 & 2 &  0 & $ 1.8 $ &  $ 0.55_{-0.33}^{+0.40} $ & $  4.17 \times 10^{42} $  & $ 6.54 \times 10^{42} $  \\ 
68	&	72	& 1.9900 & 0.23 &  0 & $ 1.90 \pm 0.15 $ &  $ 7.74_{-1.27}^{+1.25} $ & $  1.37 \times 10^{44} $  & $ 1.83 \times 10^{44} $  \\ 
76	&	66	& 0.5740 & 2 &  3 & $ 1.46 \pm 0.26 $ &  $ 6.65_{-1.15}^{+1.29} $ & $  5.88 \times 10^{42} $  & $ 1.57 \times 10^{43} $  \\ 
84	&	538	& 0.3100 & 2 &  3 & $ 1.8 $ &  $ 0.59_{-0.60}^{+4.04} $ & $  5.88 \times 10^{40} $  & $ 1.72 \times 10^{41} $  \\ 
85	&	63	& 0.5440 & 2 &  1 & $ 1.92_{-0.03}^{+0.02} $ &  $ 0.12_{-0.03}^{+0.03} $ & $  5.71 \times 10^{43} $  & $ 7.59 \times 10^{43} $  \\ 
86	&	594	& 0.7330 & 2 &  3 & $ 1.8 $ &  $ <0.12 $ & $  9.27 \times 10^{41} $  & $ 1.45 \times 10^{42} $  \\ 
92	&	60	& 1.6150 & 2 &  1 & $ 1.83_{-0.08}^{+0.09} $ &  $ 0.12_{-0.12}^{+0.33} $ & $  9.58 \times 10^{43} $  & $ 1.43 \times 10^{44} $  \\ 
93	&	97	& 0.1810 & 0 &  3 & $ 1.28_{-0.08}^{+0.13} $ &  $ <0.04 $ & $  1.47 \times 10^{41} $  & $ 5.09 \times 10^{41} $  \\ 
99	&	56	& 0.6050 & 2 &  2 & $ 1.26_{-0.12}^{+0.13} $ &  $ 1.63_{-0.29}^{+0.32} $ & $  5.58 \times 10^{42} $  & $ 2.06 \times 10^{43} $  \\ 
102	&	908	& 2.0760 & 2 &  0 & $ 1.8 $ &  $ 11.2_{-8.7}^{+11.8} $ & $  4.55 \times 10^{42} $  & $ 7.09 \times 10^{42} $  \\ 
105	&	587	& 1.8000 & 0.08 &  0 & $ 1.8 $ &  $ 6.07_{-3.42}^{+4.69} $ & $  3.70 \times 10^{42} $  & $ 5.77 \times 10^{42} $  \\ 
108	&	52	& 0.5690 & 2 &  1 & $ 1.90_{-0.09}^{+0.12} $ &  $ 0.04_{-0.04}^{+0.11} $ & $  6.14 \times 10^{42} $  & $ 8.23 \times 10^{42} $  \\ 
110	&	51	& 1.0970 & 2 &  3 & $ 1.71_{-0.22}^{+0.23} $ &  $ 22.40_{-2.90}^{+3.25} $ & $  5.49 \times 10^{43} $  & $ 1.02 \times 10^{44} $  \\ 
112	&	566	& 0.7340 & 2 &  3 & $ 1.8 $ &  $ <0.26 $ & $  4.19 \times 10^{41} $  & $ 6.57 \times 10^{41} $  \\ 
113	&	249	& 0.9640 & 2 &  4 & $ 1.8 $ &  $ 1.47_{-0.98}^{+1.32} $ & $  1.69 \times 10^{42} $  & $ 2.68 \times 10^{42} $  \\ 
115	&	525	& 0.2290 & 2 &  3 & $ 1.8 $ &  $ 0.005_{-0.005}^{+0.20} $ & $  5.29 \times 10^{40} $  & $ 1.22 \times 10^{41} $  \\ 
125	&	655	& 0.7380 & 2 &  0 & $ 1.8 $ &  $ <0.96 $ & $  1.16 \times 10^{41} $  & $ 4.39 \times 10^{41} $  \\ 
132	&	150	& 1.0900 & 2 &  4 & $ 1.8 $ &  $ 32.4_{-7.3}^{+10.0} $ & $  1.32 \times 10^{43} $  & $ 2.24 \times 10^{43} $  \\ 
133	&	46	& 1.6170 & 2 &  1 & $ 2.19_{-0.18}^{+0.20} $ &  $ 1.10_{-0.58}^{+0.50} $ & $  7.44 \times 10^{43} $  & $ 6.48 \times 10^{43} $  \\ 
134	&	42	& 0.7340 & 2 &  1 & $ 1.96_{-0.03}^{+0.01} $ &  $ 0.18_{-0.04}^{+0.03} $ & $  1.33 \times 10^{44} $  & $ 1.65 \times 10^{44} $  \\ 
138	&	651	& 0.2120 & 2 &  0 & $ 1.8 $ &  $ <0.17 $ & $  4.65 \times 10^{40} $  & $ 7.27 \times 10^{40} $  \\ 
139	&	224	& 0.7380 & 2 &  3 & $ 1.8 $ &  $ <0.22 $ & $  5.93 \times 10^{41} $  & $ 9.31\times 10^{41} $  \\ 
140	&	103	& 0.2150 & 2 &  4 & $ 1.8 $ &  $ 0.06_{-0.06}^{+0.10} $ & $  1.04 \times 10^{41} $  & $ 1.63 \times 10^{41} $  \\ 
141	&	95	& 0.0760 & 2 &  3 & $ 1.8 $ &  $ <0.010 $ & $  8.66 \times 10^{39} $  & $ 1.35 \times 10^{40} $  \\ 
142	&	116	& 0.0760 & 2 &  3 & $ 1.8 $ &  $ <0.05 $ & $  8.85 \times 10^{39} $  & $ 1.38 \times 10^{40} $  \\ 
145	&	563	& 2.2230 & 2 &  2 & $ 1.8 $ &  $ 2.5_{-2.5}^{+6.0} $ & $  3.05 \times 10^{42} $  & $ 4.74 \times 10^{42} $  \\ 
151	&	632	& 3.6200 & 0.08 &  0 & $ 1.8 $ &  $ 209_{-71}^{+156} $ & $  6.72 \times 10^{43} $  & $ 1.04 \times 10^{44} $  \\ 
154	&	247	& 0.0380 & 2 &  3 & $ 1.8 $ &  $ 1.63_{-0.77}^{+0.98} $ & $  2.49 \times 10^{39} $  & $ 4.11 \times 10^{39} $  \\ 
156	&	913	& 2.5790 & 1 &  0 & $ 1.8 $ &  $ 150 $ & $  2.05 \times 10^{43} $  & $ 3.40\times 10^{43} $  \\ 
157	&	577	& 0.5470 & 2 &  3 & $ 1.8 $ &  $ <0.14 $ & $  1.68 \times 10^{41} $  & $ 2.62 \times 10^{41} $  \\ 
162	&	31	& 1.6030 & 2 &  2 & $ 2.12_{-0.09}^{+0.08} $ &  $ 1.79_{-0.35}^{+0.33} $ & $  1.72\times 10^{44} $  & $ 1.65\times 10^{44} $  \\ 
163	&	582	& 0.2420 & 2 &  3 & $ 1.8 $ &  $ 1.70_{-1.14}^{+1.28} $ & $  2.59 \times 10^{40} $  & $ 4.05 \times 10^{40} $  \\ 
165	&	29	& 0.2980 & 2 &  0 & $ 2.03_{-0.20}^{+0.10} $ &  $ 5.31 \pm 0.56 $ & $  8.11 \times 10^{42} $  & $ 9.48\times 10^{42} $  \\ 
166	&	641	& 0.6520 & 1 &  0 & $ 1.8 $ &  $ 4.89_{-1.64}^{+1.88} $ & $  3.67 \times 10^{42} $  & $ 5.91 \times 10^{42} $  \\ 
170	&	27	& 3.0640 & 2 &  2 & $ 1.22 \pm 0.24 $ &  $ 28.0_{-8.2}^{+9.0} $ & $  5.41 \times 10^{43} $  & $ 2.07 \times 10^{44} $  \\ 
173	&	25	& 0.6250 & 0 &  4 & $ 0.32_{-0.21}^{+0.22} $ &  $ 0.57_{-0.56}^{+0.67} $ & $  5.56 \times 10^{41} $  & $ 8.71 \times 10^{42} $  \\ 
178	&	98	& 0.2790 & 2 &  2 & $ 1.8 $ &  $ <0.02 $ & $  9.22 \times 10^{40} $  & $ 1.47 \times 10^{41} $  \\ 
183	&	644	& 0.1030 & 2 &  0 & $ 1.8 $ &  $ <0.14 $ & $  8.26 \times 10^{39} $  & $ 1.29 \times 10^{40} $  \\ 
186	&	84	& 0.1030 & 2 &  4 & $ 2.04_{-0.18}^{+0.28} $ &  $ <0.07 $ & $  3.19 \times 10^{40} $  & $ 3.47 \times 10^{40} $  \\ 
190	&	18	& 0.9790 & 2 &  3 & $ 1.75 \pm 0.08 $ &  $ 1.92_{-0.22}^{+0.22} $ & $  6.59 \times 10^{43} $  & $ 1.13 \times 10^{44} $  \\ 
192	&	578	& 1.1200 & 2 &  0 & $ 1.8 $ &  $ 0.42_{-0.41}^{+2.44} $ & $  5.92 \times 10^{41} $  & $ 9.27 \times 10^{41} $  \\ 
200	&	175	& 0.5220 & 2 &  2 & $ 1.8 $ &  $ < 3.30 $ & $  1.07 \times 10^{41} $  & $ 2.56 \times 10^{41} $  \\ 
208	&	152	& 1.2800 & 0.98 &  0 & $ 1.84_{-0.36}^{+0.39} $ &  $ 19.7_{-4.1}^{+4.6} $ & $  4.40 \times 10^{43} $  & $ 6.51 \times 10^{43} $  \\ 
214	&	506	& 3.6900 & 0.57 &  0 & $ 1.8 $ &  $ 6.7_{-4.1}^{+4.6} $ & $  1.00 \times 10^{44} $  & $ 1.55 \times 10^{44} $  \\ 
218	&	608	& 0.8900 & 2 &  0 & $ 1.8 $ &  $ 150 $ & $  4.47 \times 10^{43} $  & $ 7.02 \times 10^{43} $  \\ 
225	&	650	& 0.2130 & 0.02 &  0 & $ 1.8 $ &  $ 0.47_{-0.33}^{+0.38} $ & $  9.86 \times 10^{40} $  & $ 1.54 \times 10^{41} $  \\ 
228	&	502	& 0.7320 & 2 &  0 & $ 1.8 $ &  $ 150 $ & $  5.83\times 10^{43} $  & $ 9.16 \times 10^{43} $  \\ 
230	&	501	& 1.0290 & 2 &  0 & $ 1.64_{-0.13}^{+0.11} $ &  $ 0.42_{-0.27}^{+0.25} $ & $  2.73 \times 10^{43} $  & $ 5.47 \times 10^{43} $  \\ 
\enddata
\end{deluxetable}

\clearpage

\begin{deluxetable}{rrrrrrrrr}
\tablecaption{X--ray properties of Radio sources with X--ray
matches in the complementary area covered by the E-CDFS.  X ID are from Lehmer et al. (2005) \label{table2} }
\tablehead{R ID & X ID & z & Q & Optype & $\Gamma$ & $N_H$ $10^{22}$ cm$^{-2}$ & $L_{0.5-2}$ erg s$^{-1}$ & $L_{2-10}$ erg s$^{-1}$}
\startdata
7 & 7 & 0.4940 & 0.03 &  0 & $ 1.97_{-0.10}^{+0.11} $ &  $ 0.20_{-0.11}^{+0.11} $ & $  2.53\times 10^{43} $  & $ 3.10\times 10^{43} $  \\ 
15 & 47 & 0.7890 & 0.08 &  0 & $ 1.90_{-0.24}^{+0.25} $ &  $ 7.4_{-1.0}^{+1.3} $ & $  6.08\times 10^{43} $  & $ 8.56\times 10^{43} $  \\ 
24 & 66 & 0.6900 & 0.01 &  0 & $ 1.8 $ &  $ < 0.31 $ & $  5.48\times 10^{41} $  & $ 8.66\times 10^{41} $  \\ 
26 & 82 & 0.1480 & 0.01 &  0 & $ 1.8 $ &  $ < 0.22 $ & $  1.47\times 10^{40} $  & $ 2.29\times 10^{40} $  \\ 
35 & 140 & 0.4660 & 2 &  0 & $ 1.8 $ &  $ 150 $ & $  1.41\times 10^{43} $  & $ 2.24\times 10^{43} $  \\ 
37 & 146 & 0.0570 & 0.04 &  0 & $ 1.8 $ &  $ 5.6_{-2.5}^{+3.2} $ & $  9.05\times 10^{39} $  & $ 1.41\times 10^{40} $  \\ 
52 & 194 & 0.8480 & 0.19 &  0 & $ 1.8 $ &  $ 0.31_{-0.30}^{+9.4} $ & $  4.18\times 10^{41} $  & $ 7.00\times 10^{41} $  \\ 
71 & 254 & 0.9480 & 0.04 &  0 & $ 1.8 $ &  $ < 0.15 $ & $  1.04\times 10^{43} $  & $ 1.87\times 10^{43} $  \\ 
96 & 321 & 1.5740 & 0.09 &  0 & $ 1.58_{-0.05}^{+0.02} $ &  $ < 0.05 $ & $  2.36\times 10^{44} $  & $ 6.92\times 10^{44} $  \\ 
148 & 398 & 1.9660 & 0.01 &  0 & $ 1.80_{-0.06}^{+0.07} $ &  $ < 0.15 $ & $  2.89\times 10^{44} $  & $ 7.32\times 10^{44} $  \\ 
188 & 461 & 0.1520 & 2 &  0 & $ 1.8 $ &  $ <0.143 $ & $  1.38\times 10^{40} $  & $ 3.66\times 10^{40} $  \\ 
193 & 469 & 0.5460 & 1 &  0 & $ 1.8 $ &  $ 5.6_{-1.4}^{+1.9} $ & $  2.19\times 10^{42} $  & $ 3.42\times 10^{42} $  \\ 
206 & 504 & 0.1540 & 0.01 &  0 & $ 1.8 $ &  $ < 0.04 $ & $  2.31\times 10^{40} $  & $ 3.62\times 10^{40} $  \\ 
207 & 508 & 0.6440 & 0.01 &  0 & $ 1.8 $ &  $ < 0.70 $ & $  5.13\times 10^{41} $  & $ 8.07\times 10^{41} $  \\ 
215 & 546 & 0.7340 & 0.10 &  0 & $ 1.8 $ &  $ 150 $ & $  3.34\times 10^{43} $  & $ 5.37\times 10^{43} $  \\ 
217 & 552 & 0.6230 & 2 &  0 & $ 1.8 $ &  $ 0.07_{-0.07}^{+0.28} $ & $  2.28\times 10^{42} $  & $ 3.62\times 10^{42} $  \\ 
219 & 555 & 1.0840 & 1 &  0 & $ 1.8 $ &  $ 6.4_{-5.3}^{+7.7} $ & $  1.51\times 10^{42} $  & $ 3.22\times 10^{42} $  \\ 
220 & 557 & 0.3020 & 0 &  0 & $ 1.8 $ &  $ <0.33 $ & $  3.15\times 10^{40} $  & $ 1.04\times 10^{41} $  \\ 
232 & 609 & 1.0590 & 0.12 &  0 & $ 1.8 $ &  $ 366_{-113}^{+228} $ & $  1.35\times 10^{44} $  & $ 2.44\times 10^{44} $  \\ 
240 & 634 & 0.0860 & 2 &  0 & $ 1.8 $ &  $ 0.18_{-0.18}^{+0.34} $ & $  1.20\times 10^{40} $  & $ 1.87\times 10^{40} $  \\ 
243 & 639 & 0.1480 & 2 &  0 & $ 1.8 $ &  $ <0.29 $ & $  6.55\times 10^{39} $  & $ 1.02\times 10^{40} $  \\ 
246 & 646 & 1.1280 & 0.03 &  0 & $ 1.8 $ &  $ < 2.54 $ & $  9.82\times 10^{41} $  & $ 1.82\times 10^{42} $  \\ 
249 & 657 & 0.1390 & 0.01 &  0 & $ 1.8 $ &  $ <0.12 $ & $  1.99\times 10^{40} $  & $ 3.76\times 10^{40} $  \\ 
250 & 664 & 0.1260 & 2 &  0 & $ 1.8 $ &  $ < 0.01 $ & $  9.08\times 10^{40} $  & $ 1.42\times 10^{41} $  \\ 
251 & 669 & 0.1290 & 2 &  0 & $ 3.98_{-0.59}^{+0.86} $ &  $ 0.13_{-0.13}^{+0.20} $ & $  4.96\times 10^{41} $  & $ 3.33\times 10^{40} $  \\ 
252 & 674 & 1.1510 & 0.15 &  0 & $ 1.8 $ &  $ 0.17_{-0.17}^{+1.29} $ & $  2.76\times 10^{42} $  & $ 5.18\times 10^{42} $  \\ 
259 & 738 & 0.8600 & 2 &  0 & $ 1.94_{-0.56}^{+0.58} $ &  $ 12.1_{-3.7}^{+4.6} $ & $  3.69\times 10^{43} $  & $ 5.06\times 10^{43} $  \\ 
\enddata
\end{deluxetable}

\clearpage

\begin{deluxetable}{rrrrrr}
\tablecaption{X--ray properties of Radio sources with X--ray matches and no
  redshift information in the 1Ms CDFS field or in the complementary
  area covered by the E-CDFS.  X ID are from Giacconi et al. (2002) or
  Lehmer et al. (2005)\label{table3} } 
\tablehead{R ID & X ID & Soft Cts & Hard Cts & Soft flux erg s$^{-1}$cm$^{-2}$ & Hard Flux erg s$^{-1}$cm$^{-2}$}
\startdata
14 & E46 & 144.2 & 51.8 &   $ 0.39\times 10^{-14} $  & $ 0.51\times 10^{-14} $  \\ 
18 & E51 & 24.4 & 16.9 &   $ 0.63\times 10^{-15} $  & $ 0.19\times 10^{-14} $  \\ 
33 & E136 & 49.7 & 101.2 &   $ 0.11\times 10^{-14} $  & $ 0.13\times 10^{-13} $  \\ 
49 & E188 & 9.2 & 47.0 &   $ 0.35\times 10^{-15} $  & $ 0.13\times 10^{-13} $  \\ 
56 & E205 & 7.9 & 8.6 &   $ 0.19\times 10^{-15} $  & $ <0.86\times 10^{-15} $  \\ 
73 & 70 & 115.6 & 344.6 &   $ 0.71\times 10^{-15} $  & $ 0.12\times 10^{-13} $  \\ 
80 & E289 & 174.2 & 111.0 &   $ 0.64\times 10^{-14} $  & $ 0.17\times 10^{-13} $  \\ 
87 & 537 & 16.4 & 18.1 &   $ 0.98\times 10^{-16} $  & $ 0.62\times 10^{-15} $  \\ 
122 & 570 & 23.8 & 9.3 &   $ 0.15\times 10^{-15} $  & $ 0.35\times 10^{-15} $  \\ 
176 & E437 & 78.4 & 157.7 &   $ 0.20\times 10^{-14} $  & $ 0.23\times 10^{-13} $  \\ 
211 & E535 & 9.5 & 16.8 &   $ <0.22\times 10^{-15} $  & $ 0.22\times 10^{-14} $  \\ 
231 & E599 & 9.5 & 15.7 &   $ <0.24\times 10^{-15} $  & $ 0.22\times 10^{-14} $  \\ 
262 & E743 & 16.0 & 18.2 &   $ <0.42\times 10^{-15} $  & $ <0.20\times 10^{-14} $  \\ 
\enddata
\end{deluxetable}

\clearpage

\begin{deluxetable}{rrrrrrrrr}
\tabletypesize{\scriptsize}
\tablecolumns{9}
\tablewidth{0pc}
\tablecaption{ X--ray photometric properties of radio sources without X--ray matches.  Upper limits are at 3$\sigma$.\label{table4} }
\tablehead{R ID & Soft Cts & Hard Cts & Soft S/N & Hard S/N & HR & $S_{0.5-2}$  erg cm$^{-2}$ s$^{-1}$  & $S_{2-10}$  erg cm$^{-2}$ s$^{-1}$  & z}
\startdata
    2 & $  <   21.60 $ & $  <    9.30 $ & $     0.0 $ & $     0.0 $ &   -   & $  <   0.64 \times 10^{-15} $ & $  <   0.12 \times 10^{-14} $ & $   - $ 	\\
    3 & $  <   15.30 $ & $  <   28.70 $ & $     0.0 $ & $     0.0 $ &   -   & $  <   0.51 \times 10^{-15} $ & $  <   0.43 \times 10^{-14} $ & $   - $ 	\\
    4 & $   10.90 \pm    5.10 $ & $ <    10.00 $ & $     2.3 $ & $     0.0 $ & $   -1.00 $ & $   0.32 \times 10^{-15} $ & $ <  0.13 \times 10^{-14} $ & $   - $ 	\\
    5 & $  <   17.20 $ & $  <   23.50 $ & $     0.0 $ & $     0.0 $ &   -   & $  <   0.50 \times 10^{-15} $ & $  <   0.36 \times 10^{-14} $ & $   - $ 	\\
    6 & $  <   20.10 $ & $  <   18.90 $ & $     0.0 $ & $     0.0 $ &   -   & $  <   0.59 \times 10^{-15} $ & $  <   0.29 \times 10^{-14} $ & $   - $ 	\\
    8 & $  <   12.20 $ & $  <   16.80 $ & $     0.0 $ & $     0.0 $ &   -   & $  <   0.36 \times 10^{-15} $ & $  <   0.22 \times 10^{-14} $ & $    0.86 $ 	\\
    9 & $  <   11.70 $ & $  <   14.10 $ & $     0.0 $ & $     0.0 $ &   -   & $  <   0.33 \times 10^{-15} $ & $  <   0.21 \times 10^{-14} $ & $    0.67 $ 	\\
   10 & $  <   20.40 $ & $  <   21.00 $ & $     0.0 $ & $     0.0 $ &   -   & $  <   0.65 \times 10^{-15} $ & $  <   0.30 \times 10^{-14} $ & $   - $ 	\\
   11 & $  <    6.90 $ & $  <   21.30 $ & $     0.0 $ & $     0.0 $ &   -   & $  <   0.20 \times 10^{-15} $ & $  <   0.26 \times 10^{-14} $ & $    0.90 $ 	\\
   12 & $  <   13.30 $ & $  <   15.60 $ & $     0.0 $ & $     0.0 $ &   -   & $  <   0.39 \times 10^{-15} $ & $  <   0.20 \times 10^{-14} $ & $    0.55 $ 	\\
   13 & $  <   23.70 $ & $  <   10.80 $ & $     0.0 $ & $     0.0 $ &   -   & $  <   0.73 \times 10^{-15} $ & $  <   0.15 \times 10^{-14} $ & $    1.00 $ 	\\
   16 & $  <   16.00 $ & $  <   18.10 $ & $     0.0 $ & $     0.0 $ &   -   & $  <   0.45 \times 10^{-15} $ & $  <   0.26 \times 10^{-14} $ & $   - $ 	\\
   17 & $  <   11.50 $ & $  <   11.20 $ & $     0.0 $ & $     0.0 $ &   -   & $  <   0.31 \times 10^{-15} $ & $  <   0.13 \times 10^{-14} $ & $    0.55 $ 	\\
   19 & $  <    6.60 $ & $  <    6.00 $ & $     0.0 $ & $     0.0 $ &   -   & $  <   0.19 \times 10^{-15} $ & $  <   0.73 \times 10^{-15} $ & $   - $ 	\\
   20 & $  <   13.60 $ & $  <   24.40 $ & $     0.0 $ & $     0.0 $ &   -   & $  <   0.43 \times 10^{-15} $ & $  <   0.41 \times 10^{-14} $ & $    0.85 $ 	\\
   22 & $  <   11.30 $ & $  <   12.70 $ & $     0.0 $ & $     0.0 $ &   -   & $  <   0.30 \times 10^{-15} $ & $  <   0.17 \times 10^{-14} $ & $   - $ 	\\
   23 & $  <    6.90 $ & $  <    9.40 $ & $     0.0 $ & $     0.0 $ &   -   & $  <   0.19 \times 10^{-15} $ & $  <   0.11 \times 10^{-14} $ & $   - $ 	\\
   25 & $  <    7.40 $ & $  <    3.20 $ & $     0.0 $ & $     0.0 $ &   -   & $  <   0.21 \times 10^{-15} $ & $  <   0.46 \times 10^{-15} $ & $    1.06 $ 	\\
   27 & $  <    8.10 $ & $  <    6.70 $ & $     0.0 $ & $     0.0 $ &   -   & $  <   0.21 \times 10^{-15} $ & $  <   0.75 \times 10^{-15} $ & $    1.08 $ 	\\
   28 & $  <    7.30 $ & $  <   14.40 $ & $     0.0 $ & $     0.0 $ &   -   & $  <   0.20 \times 10^{-15} $ & $  <   0.21 \times 10^{-14} $ & $    0.36 $ 	\\
   29 & $  <    7.60 $ & $  <   11.80 $ & $     0.0 $ & $     0.0 $ &   -   & $  <   0.20 \times 10^{-15} $ & $  <   0.15 \times 10^{-14} $ & $   - $ 	\\
   30 & $  <    9.80 $ & $  <    3.80 $ & $     0.0 $ & $     0.0 $ &   -   & $  <   0.26 \times 10^{-15} $ & $  <   0.42 \times 10^{-15} $ & $   - $ 	\\
   31 & $  <    6.80 $ & $  <    4.30 $ & $     0.0 $ & $     0.0 $ &   -   & $  <   0.23 \times 10^{-15} $ & $  <   0.62 \times 10^{-15} $ & $   - $ 	\\
   32 & $  <    6.20 $ & $  <    6.30 $ & $     0.0 $ & $     0.0 $ &   -   & $  <   0.17 \times 10^{-15} $ & $  <   0.74 \times 10^{-15} $ & $    0.69 $ 	\\
   34 & $  <   50.10 $ & $  <   69.50 $ & $     0.0 $ & $     0.0 $ &   -   & $  <   0.39 \times 10^{-15} $ & $  <   0.32 \times 10^{-14} $ & $    0.26 $ 	\\
   36 & $  <   10.20 $ & $  <   16.30 $ & $     0.0 $ & $     0.0 $ &   -   & $  <   0.27 \times 10^{-15} $ & $  <   0.22 \times 10^{-14} $ & $    0.11 $ 	\\
   38 & $  <    1.20 $ & $  <    8.40 $ & $     0.0 $ & $     0.0 $ &   -   & $  <   0.36 \times 10^{-16} $ & $  <   0.10 \times 10^{-14} $ & $    0.83 $ 	\\
   39 & $  <   38.80 $ & $  <   59.00 $ & $     0.0 $ & $     0.0 $ &   -   & $  <   0.29 \times 10^{-15} $ & $  <   0.26 \times 10^{-14} $ & $    0.68 $ 	\\
   40 & $  <   12.20 $ & $  <   14.30 $ & $     0.0 $ & $     0.0 $ &   -   & $  <   0.32 \times 10^{-15} $ & $  <   0.19 \times 10^{-14} $ & $   - $ 	\\
   41 & $  <    8.80 $ & $  <    4.00 $ & $     0.0 $ & $     0.0 $ &   -   & $  <   0.24 \times 10^{-15} $ & $  <   0.55 \times 10^{-15} $ & $    0.36 $ 	\\
   42 & $  <    1.90 $ & $  <    4.80 $ & $     0.0 $ & $     0.0 $ &   -   & $  <   0.68 \times 10^{-16} $ & $  <   0.73 \times 10^{-15} $ & $   - $ 	\\
   43 & $   21.90 \pm   10.50 $ & $ <    47.00 $ & $     2.3 $ & $     0.0 $ & $   -1.00 $ & $   0.16 \times 10^{-15} $ & $ <  0.21 \times 10^{-14} $ & $    0.54 $ 	\\
   44 & $  <   10.20 $ & $  <    6.60 $ & $     0.0 $ & $     0.0 $ &   -   & $  <   0.27 \times 10^{-15} $ & $  <   0.88 \times 10^{-15} $ & $   - $ 	\\
   45 & $    6.30 \pm    2.90 $ & $ <     8.70 $ & $     2.2 $ & $     0.0 $ & $   -1.00 $ & $   0.20 \times 10^{-15} $ & $ <  0.12 \times 10^{-14} $ & $    0.62 $ 	\\
   46 & $  <   51.80 $ & $  <   44.00 $ & $     0.0 $ & $     0.0 $ &   -   & $  <   0.37 \times 10^{-15} $ & $  <   0.19 \times 10^{-14} $ & $   - $ 	\\
   47 & $  <    6.20 $ & $  <    9.50 $ & $     0.0 $ & $     0.0 $ &   -   & $  <   0.20 \times 10^{-15} $ & $  <   0.16 \times 10^{-14} $ & $   - $ 	\\
   48 & $  <   12.00 $ & $  <    8.90 $ & $     0.0 $ & $     0.0 $ &   -   & $  <   0.74 \times 10^{-15} $ & $  <   0.23 \times 10^{-14} $ & $    1.13 $ 	\\
   50 & $  <    4.80 $ & $  <    3.80 $ & $     0.0 $ & $     0.0 $ &   -   & $  <   0.33 \times 10^{-15} $ & $  <   0.11 \times 10^{-14} $ & $    1.79 $ 	\\
   51 & $  <   32.40 $ & $   31.70 \pm   11.30 $ & $     0.0 $ & $     3.1 $ & $    1.00 $ & $  <   0.25 \times 10^{-15} $ & $   0.15 \times 10^{-14} $ & $   - $	\\
   53 & $  <   16.70 $ & $  <   40.70 $ & $     0.0 $ & $     0.0 $ &   -   & $  <   0.11 \times 10^{-15} $ & $  <   0.16 \times 10^{-14} $ & $   - $ 	\\
   55 & $  <    7.60 $ & $  <    8.60 $ & $     0.0 $ & $     0.0 $ &   -   & $  <   0.20 \times 10^{-15} $ & $  <   0.96 \times 10^{-15} $ & $    1.50 $ 	\\
   57 & $  <    5.40 $ & $  <    3.90 $ & $     0.0 $ & $     0.0 $ &   -   & $  <   0.14 \times 10^{-15} $ & $  <   0.43 \times 10^{-15} $ & $   - $ 	\\
   58 & $  <   12.10 $ & $  <   10.20 $ & $     0.0 $ & $     0.0 $ &   -   & $  <   0.33 \times 10^{-15} $ & $  <   0.14 \times 10^{-14} $ & $    0.96 $ 	\\
   59 & $  <   24.00 $ & $   20.50 \pm   10.60 $ & $     0.0 $ & $     2.1 $ & $    1.00 $ & $  <   0.16 \times 10^{-15} $ & $   0.78 \times 10^{-15} $ & $    0.71 $	\\
   60 & $  <   47.00 $ & $  <   55.00 $ & $     0.0 $ & $     0.0 $ &   -   & $  <   0.31 \times 10^{-15} $ & $  <   0.21 \times 10^{-14} $ & $   - $ 	\\
   61 & $  <   13.80 $ & $  <   15.90 $ & $     0.0 $ & $     0.0 $ &   -   & $  <   0.30 \times 10^{-15} $ & $  <   0.20 \times 10^{-14} $ & $    0.76 $ 	\\
   62 & $  <    1.30 $ & $  <    8.40 $ & $     0.0 $ & $     0.0 $ &   -   & $  <   0.35 \times 10^{-16} $ & $  <   0.94 \times 10^{-15} $ & $   - $ 	\\
   63 & $  <   19.50 $ & $  <   40.10 $ & $     0.0 $ & $     0.0 $ &   -   & $  <   0.12 \times 10^{-15} $ & $  <   0.15 \times 10^{-14} $ & $    0.77 $ 	\\
   64 & $  <   26.70 $ & $  <   13.30 $ & $     0.0 $ & $     0.0 $ &   -   & $  <   0.17 \times 10^{-15} $ & $  <   0.49 \times 10^{-15} $ & $   - $ 	\\
   65 & $  <    1.40 $ & $  <   -0.10 $ & $     0.0 $ & $     0.0 $ &   -   & $  <   0.36 \times 10^{-16} $ & $  <  -0.13 \times 10^{-16} $ & $   - $ 	\\
   67 & $  <    6.20 $ & $  <    8.80 $ & $     0.0 $ & $     0.0 $ &   -   & $  <   0.16 \times 10^{-15} $ & $  <   0.98 \times 10^{-15} $ & $   - $ 	\\
   69 & $  <   13.70 $ & $  <   10.10 $ & $     0.0 $ & $     0.0 $ &   -   & $  <   0.27 \times 10^{-15} $ & $  <   0.12 \times 10^{-14} $ & $    0.81 $ 	\\
   70 & $  <   16.10 $ & $  <   10.50 $ & $     0.0 $ & $     0.0 $ &   -   & $  <   0.11 \times 10^{-15} $ & $  <   0.43 \times 10^{-15} $ & $   - $ 	\\
   72 & $  <   34.30 $ & $  <   39.40 $ & $     0.0 $ & $     0.0 $ &   -   & $  <   0.21 \times 10^{-15} $ & $  <   0.14 \times 10^{-14} $ & $   - $ 	\\
   74 & $  <    1.50 $ & $  <   10.50 $ & $     0.0 $ & $     0.0 $ &   -   & $  <   0.39 \times 10^{-16} $ & $  <   0.14 \times 10^{-14} $ & $   - $ 	\\
   75 & $  <   20.20 $ & $  <   23.20 $ & $     0.0 $ & $     0.0 $ &   -   & $  <   0.13 \times 10^{-15} $ & $  <   0.86 \times 10^{-15} $ & $   - $ 	\\
   77 & $  <    4.10 $ & $  <    7.00 $ & $     0.0 $ & $     0.0 $ &   -   & $  <   0.12 \times 10^{-15} $ & $  <   0.10 \times 10^{-14} $ & $   - $ 	\\
   78 & $  <    8.30 $ & $  <    9.00 $ & $     0.0 $ & $     0.0 $ &   -   & $  <   0.22 \times 10^{-15} $ & $  <   0.12 \times 10^{-14} $ & $   - $ 	\\
   79 & $  <   19.20 $ & $  <    8.80 $ & $     0.0 $ & $     0.0 $ &   -   & $  <   0.12 \times 10^{-15} $ & $  <   0.31 \times 10^{-15} $ & $   - $ 	\\
   81 & $  <   11.50 $ & $  <    5.40 $ & $     0.0 $ & $     0.0 $ &   -   & $  <   0.30 \times 10^{-15} $ & $  <   0.73 \times 10^{-15} $ & $    0.50 $ 	\\
   82 & $  <   16.00 $ & $  <   19.90 $ & $     0.0 $ & $     0.0 $ &   -   & $  <   0.97 \times 10^{-16} $ & $  <   0.70 \times 10^{-15} $ & $    1.02 $ 	\\
   83 & $  <   18.80 $ & $  <   17.40 $ & $     0.0 $ & $     0.0 $ &   -   & $  <   0.12 \times 10^{-15} $ & $  <   0.66 \times 10^{-15} $ & $    0.03 $ 	\\
   88 & $  <    4.40 $ & $  <    8.00 $ & $     0.0 $ & $     0.0 $ &   -   & $  <   0.12 \times 10^{-15} $ & $  <   0.11 \times 10^{-14} $ & $    0.65 $ 	\\
   89 & $  <   11.10 $ & $  <   24.30 $ & $     0.0 $ & $     0.0 $ &   -   & $  <   0.66 \times 10^{-16} $ & $  <   0.83 \times 10^{-15} $ & $    0.65 $ 	\\
   90 & $  <   12.10 $ & $  <   17.60 $ & $     0.0 $ & $     0.0 $ &   -   & $  <   0.33 \times 10^{-15} $ & $  <   0.24 \times 10^{-14} $ & $    0.65 $ 	\\
   91 & $  <   10.30 $ & $  <   14.80 $ & $     0.0 $ & $     0.0 $ &   -   & $  <   0.64 \times 10^{-16} $ & $  <   0.53 \times 10^{-15} $ & $    1.61 $ 	\\
   94 & $  <   25.50 $ & $  <   12.60 $ & $     0.0 $ & $     0.0 $ &   -   & $  <   0.15 \times 10^{-15} $ & $  <   0.43 \times 10^{-15} $ & $    0.55 $ 	\\
   95 & $  <   15.00 $ & $  <   23.40 $ & $     0.0 $ & $     0.0 $ &   -   & $  <   0.10 \times 10^{-15} $ & $  <   0.92 \times 10^{-15} $ & $    0.58 $ 	\\
   97 & $  <   29.40 $ & $  <   13.60 $ & $     0.0 $ & $     0.0 $ &   -   & $  <   0.18 \times 10^{-15} $ & $  <   0.48 \times 10^{-15} $ & $   - $ 	\\
   98 & $  <    8.90 $ & $  <    5.20 $ & $     0.0 $ & $     0.0 $ &   -   & $  <   0.54 \times 10^{-15} $ & $  <   0.16 \times 10^{-14} $ & $    1.05 $ 	\\
  100 & $  <   18.80 $ & $  <   30.10 $ & $     0.0 $ & $     0.0 $ &   -   & $  <   0.22 \times 10^{-15} $ & $  <   0.21 \times 10^{-14} $ & $   - $ 	\\
  101 & $  <   19.10 $ & $  <   16.00 $ & $     0.0 $ & $     0.0 $ &   -   & $  <   0.11 \times 10^{-15} $ & $  <   0.55 \times 10^{-15} $ & $    0.05 $ 	\\
  103 & $  <   35.80 $ & $  <   32.00 $ & $     0.0 $ & $     0.0 $ &   -   & $  <   0.33 \times 10^{-15} $ & $  <   0.18 \times 10^{-14} $ & $    0.73 $ 	\\
  104 & $  <   12.40 $ & $  <    3.20 $ & $     0.0 $ & $     0.0 $ &   -   & $  <   0.33 \times 10^{-15} $ & $  <   0.37 \times 10^{-15} $ & $   - $ 	\\
  106 & $  <   23.40 $ & $   17.10 \pm    7.20 $ & $     0.0 $ & $     2.6 $ & $    1.00 $ & $  <   0.67 \times 10^{-15} $ & $   0.21 \times 10^{-14} $ & $    0.76 $	\\
  107 & $  <    4.90 $ & $  <   10.00 $ & $     0.0 $ & $     0.0 $ &   -   & $  <   0.14 \times 10^{-15} $ & $  <   0.14 \times 10^{-14} $ & $   - $ 	\\
  109 & $    8.70 \pm    4.00 $ & $ <     0.40 $ & $     2.2 $ & $     0.0 $ & $   -1.00 $ & $   0.24 \times 10^{-15} $ & $ <  0.57 \times 10^{-16} $ & $    0.14 $ 	\\
  111 & $  <   20.40 $ & $  <   21.70 $ & $     0.0 $ & $     0.0 $ &   -   & $  <   0.12 \times 10^{-15} $ & $  <   0.76 \times 10^{-15} $ & $    0.12 $ 	\\
  114 & $  <   17.30 $ & $  <   20.60 $ & $     0.0 $ & $     0.0 $ &   -   & $  <   0.13 \times 10^{-15} $ & $  <   0.88 \times 10^{-15} $ & $    1.00 $ 	\\
  116 & $  <   13.90 $ & $  <   27.10 $ & $     0.0 $ & $     0.0 $ &   -   & $  <   0.13 \times 10^{-15} $ & $  <   0.15 \times 10^{-14} $ & $    0.70 $ 	\\
  117 & $  <   18.70 $ & $  <   21.10 $ & $     0.0 $ & $     0.0 $ &   -   & $  <   0.15 \times 10^{-15} $ & $  <   0.96 \times 10^{-15} $ & $    0.34 $ 	\\
  118 & $  <   18.90 $ & $  <   16.60 $ & $     0.0 $ & $     0.0 $ &   -   & $  <   0.11 \times 10^{-15} $ & $  <   0.57 \times 10^{-15} $ & $    0.67 $ 	\\
  119 & $  <   18.60 $ & $  <    8.00 $ & $     0.0 $ & $     0.0 $ &   -   & $  <   0.52 \times 10^{-15} $ & $  <   0.97 \times 10^{-15} $ & $    0.71 $ 	\\
  120 & $   13.00 \pm    5.70 $ & $ <    21.60 $ & $     2.3 $ & $     0.0 $ & $   -1.00 $ & $   0.77 \times 10^{-16} $ & $ <  0.74 \times 10^{-15} $ & $    0.52 $ 	\\
  121 & $  <   20.90 $ & $  <   18.90 $ & $     0.0 $ & $     0.0 $ &   -   & $  <   0.13 \times 10^{-15} $ & $  <   0.69 \times 10^{-15} $ & $    2.12 $ 	\\
  123 & $  <   23.50 $ & $  <   24.10 $ & $     0.0 $ & $     0.0 $ &   -   & $  <   0.14 \times 10^{-15} $ & $  <   0.82 \times 10^{-15} $ & $    0.73 $ 	\\
  124 & $  <   15.80 $ & $  <   19.90 $ & $     0.0 $ & $     0.0 $ &   -   & $  <   0.45 \times 10^{-15} $ & $  <   0.30 \times 10^{-14} $ & $    0.13 $ 	\\
  126 & $  <   21.80 $ & $  <   18.90 $ & $     0.0 $ & $     0.0 $ &   -   & $  <   0.14 \times 10^{-15} $ & $  <   0.70 \times 10^{-15} $ & $    1.71 $ 	\\
  127 & $  <   11.60 $ & $  <   22.60 $ & $     0.0 $ & $     0.0 $ &   -   & $  <   0.32 \times 10^{-15} $ & $  <   0.27 \times 10^{-14} $ & $    0.12 $ 	\\
  128 & $  <   16.10 $ & $  <   24.00 $ & $     0.0 $ & $     0.0 $ &   -   & $  <   0.46 \times 10^{-15} $ & $  <   0.36 \times 10^{-14} $ & $    0.12 $ 	\\
  129 & $  <   12.00 $ & $  <   21.30 $ & $     0.0 $ & $     0.0 $ &   -   & $  <   0.35 \times 10^{-15} $ & $  <   0.33 \times 10^{-14} $ & $    0.54 $ 	\\
  130 & $   16.90 \pm    6.20 $ & $ <    29.60 $ & $     2.8 $ & $     0.0 $ & $   -1.00 $ & $   0.11 \times 10^{-15} $ & $ <  0.11 \times 10^{-14} $ & $    0.13 $ 	\\
  131 & $  <   34.80 $ & $  <   16.90 $ & $     0.0 $ & $     0.0 $ &   -   & $  <   0.68 \times 10^{-15} $ & $  <   0.20 \times 10^{-14} $ & $    0.61 $ 	\\
  135 & $   12.90 \pm    5.40 $ & $ <    23.40 $ & $     2.6 $ & $     0.0 $ & $   -1.00 $ & $   0.36 \times 10^{-15} $ & $ <  0.28 \times 10^{-14} $ & $    1.12 $ 	\\
  136 & $  <   22.90 $ & $  <   11.40 $ & $     0.0 $ & $     0.0 $ &   -   & $  <   0.15 \times 10^{-15} $ & $  <   0.43 \times 10^{-15} $ & $    0.25 $ 	\\
  137 & $  <   29.80 $ & $  <   43.50 $ & $     0.0 $ & $     0.0 $ &   -   & $  <   0.58 \times 10^{-15} $ & $  <   0.50 \times 10^{-14} $ & $    0.39 $ 	\\
  143 & $   32.20 \pm   13.00 $ & $ <    57.40 $ & $     3.0 $ & $     0.0 $ & $   -1.00 $ & $   0.40 \times 10^{-15} $ & $ <  0.43 \times 10^{-14} $ & $    0.13 $ 	\\
  144 & $  <   23.10 $ & $  <   26.90 $ & $     0.0 $ & $     0.0 $ &   -   & $  <   0.24 \times 10^{-15} $ & $  <   0.17 \times 10^{-14} $ & $   - $ 	\\
  146 & $   15.70 \pm    5.40 $ & $ <    18.90 $ & $     3.0 $ & $     0.0 $ & $   -1.00 $ & $   0.91 \times 10^{-16} $ & $ <  0.63 \times 10^{-15} $ & $    1.56 $ 	\\
  147 & $  <   19.30 $ & $  <   39.40 $ & $     0.0 $ & $     0.0 $ &   -   & $  <   0.54 \times 10^{-15} $ & $  <   0.61 \times 10^{-14} $ & $   - $ 	\\
  149 & $  <   16.20 $ & $  <   22.90 $ & $     0.0 $ & $     0.0 $ &   -   & $  <   0.95 \times 10^{-16} $ & $  <   0.79 \times 10^{-15} $ & $    0.36 $ 	\\
  150 & $  <   20.80 $ & $  <   19.80 $ & $     0.0 $ & $     0.0 $ &   -   & $  <   0.14 \times 10^{-15} $ & $  <   0.75 \times 10^{-15} $ & $    0.67 $ 	\\
  152 & $  <   28.70 $ & $  <   32.70 $ & $     0.0 $ & $     0.0 $ &   -   & $  <   0.35 \times 10^{-15} $ & $  <   0.24 \times 10^{-14} $ & $    0.21 $ 	\\
  153 & $  <   12.10 $ & $  <   20.00 $ & $     0.0 $ & $     0.0 $ &   -   & $  <   0.32 \times 10^{-15} $ & $  <   0.28 \times 10^{-14} $ & $    0.68 $ 	\\
  155 & $  <   15.40 $ & $  <   33.30 $ & $     0.0 $ & $     0.0 $ &   -   & $  <   0.97 \times 10^{-16} $ & $  <   0.12 \times 10^{-14} $ & $   - $ 	\\
  158 & $  <   18.30 $ & $  <   30.90 $ & $     0.0 $ & $     0.0 $ &   -   & $  <   0.58 \times 10^{-15} $ & $  <   0.43 \times 10^{-14} $ & $   - $ 	\\
  159 & $  <   42.50 $ & $  <   58.70 $ & $     0.0 $ & $     0.0 $ &   -   & $  <   0.38 \times 10^{-15} $ & $  <   0.31 \times 10^{-14} $ & $    0.13 $ 	\\
  160 & $  <   19.80 $ & $   18.80 \pm    6.70 $ & $     0.0 $ & $     2.9 $ & $    1.00 $ & $  <   0.12 \times 10^{-15} $ & $   0.67 \times 10^{-15} $ & $    0.52 $	\\
  161 & $  <   10.80 $ & $  <    9.30 $ & $     0.0 $ & $     0.0 $ &   -   & $  <   0.63 \times 10^{-16} $ & $  <   0.31 \times 10^{-15} $ & $    1.62 $ 	\\
  164 & $   11.20 \pm    5.40 $ & $ <    19.10 $ & $     2.2 $ & $     0.0 $ & $   -1.00 $ & $   0.72 \times 10^{-16} $ & $ <  0.72 \times 10^{-15} $ & $    0.59 $ 	\\
  167 & $  <    7.20 $ & $  <   11.90 $ & $     0.0 $ & $     0.0 $ &   -   & $  <   0.44 \times 10^{-16} $ & $  <   0.42 \times 10^{-15} $ & $    0.61 $ 	\\
  168 & $  <   15.80 $ & $  <   18.00 $ & $     0.0 $ & $     0.0 $ &   -   & $  <   0.98 \times 10^{-16} $ & $  <   0.65 \times 10^{-15} $ & $    1.29 $ 	\\
  169 & $   10.40 \pm    4.80 $ & $ <    14.30 $ & $     2.2 $ & $     0.0 $ & $   -1.00 $ & $   0.64 \times 10^{-16} $ & $ <  0.51 \times 10^{-15} $ & $    0.69 $ 	\\
  171 & $  <   21.80 $ & $  <   14.80 $ & $     0.0 $ & $     0.0 $ &   -   & $  <   0.59 \times 10^{-15} $ & $  <   0.21 \times 10^{-14} $ & $   - $ 	\\
  172 & $  <   16.70 $ & $  <   32.80 $ & $     0.0 $ & $     0.0 $ &   -   & $  <   0.11 \times 10^{-15} $ & $  <   0.12 \times 10^{-14} $ & $    0.98 $ 	\\
  174 & $  <    5.90 $ & $  <   10.10 $ & $     0.0 $ & $     0.0 $ &   -   & $  <   0.16 \times 10^{-15} $ & $  <   0.14 \times 10^{-14} $ & $    0.61 $ 	\\
  175 & $  <   14.10 $ & $  <    5.90 $ & $     0.0 $ & $     0.0 $ &   -   & $  <   0.40 \times 10^{-15} $ & $  <   0.70 \times 10^{-15} $ & $    0.15 $ 	\\
  177 & $  <   30.90 $ & $  <   31.30 $ & $     0.0 $ & $     0.0 $ &   -   & $  <   0.20 \times 10^{-15} $ & $  <   0.12 \times 10^{-14} $ & $    0.95 $ 	\\
  179 & $  <   35.30 $ & $  <   34.30 $ & $     0.0 $ & $     0.0 $ &   -   & $  <   0.30 \times 10^{-15} $ & $  <   0.17 \times 10^{-14} $ & $    1.09 $ 	\\
  180 & $  <   24.80 $ & $  <   14.90 $ & $     0.0 $ & $     0.0 $ &   -   & $  <   0.15 \times 10^{-15} $ & $  <   0.51 \times 10^{-15} $ & $    0.19 $ 	\\
  181 & $   24.60 \pm    7.90 $ & $ <    27.70 $ & $     3.3 $ & $     0.0 $ & $   -1.00 $ & $   0.15 \times 10^{-15} $ & $ <  0.10 \times 10^{-14} $ & $    0.46 $ 	\\
  182 & $  <   15.10 $ & $  <   23.70 $ & $     0.0 $ & $     0.0 $ &   -   & $  <   0.41 \times 10^{-15} $ & $  <   0.35 \times 10^{-14} $ & $   - $ 	\\
  184 & $  <   15.80 $ & $  <   17.10 $ & $     0.0 $ & $     0.0 $ &   -   & $  <   0.98 \times 10^{-16} $ & $  <   0.61 \times 10^{-15} $ & $    2.31 $ 	\\
  185 & $   18.60 \pm    6.70 $ & $ <    21.20 $ & $     2.9 $ & $     0.0 $ & $   -1.00 $ & $   0.12 \times 10^{-15} $ & $ <  0.77 \times 10^{-15} $ & $    1.99 $ 	\\
  187 & $   11.70 \pm    4.80 $ & $ <    20.20 $ & $     2.5 $ & $     0.0 $ & $   -1.00 $ & $   0.83 \times 10^{-16} $ & $ <  0.83 \times 10^{-15} $ & $    2.29 $ 	\\
  189 & $   18.00 \pm    8.40 $ & $ <    35.70 $ & $     2.3 $ & $     0.0 $ & $   -1.00 $ & $   0.12 \times 10^{-15} $ & $ <  0.14 \times 10^{-14} $ & $    1.30 $ 	\\
  191 & $  <   42.50 $ & $   36.50 \pm   13.30 $ & $     0.0 $ & $     3.1 $ & $    1.00 $ & $  <   0.28 \times 10^{-15} $ & $   0.14 \times 10^{-14} $ & $    0.35 $	\\
  194 & $  <    5.30 $ & $  <    4.50 $ & $     0.0 $ & $     0.0 $ &   -   & $  <   0.14 \times 10^{-15} $ & $  <   0.60 \times 10^{-15} $ & $   - $ 	\\
  195 & $  <   34.50 $ & $  <   28.80 $ & $     0.0 $ & $     0.0 $ &   -   & $  <   0.22 \times 10^{-15} $ & $  <   0.11 \times 10^{-14} $ & $    0.98 $ 	\\
  196 & $  <   11.50 $ & $  <   11.40 $ & $     0.0 $ & $     0.0 $ &   -   & $  <   0.31 \times 10^{-15} $ & $  <   0.13 \times 10^{-14} $ & $    0.25 $ 	\\
  197 & $   14.30 \pm    4.80 $ & $ <    10.00 $ & $     3.1 $ & $     0.0 $ & $   -1.00 $ & $   0.39 \times 10^{-15} $ & $ <  0.12 \times 10^{-14} $ & $    0.25 $ 	\\
  198 & $  <    6.00 $ & $  <    6.60 $ & $     0.0 $ & $     0.0 $ &   -   & $  <   0.16 \times 10^{-15} $ & $  <   0.75 \times 10^{-15} $ & $    0.78 $ 	\\
  199 & $  <   12.10 $ & $  <   12.80 $ & $     0.0 $ & $     0.0 $ &   -   & $  <   0.32 \times 10^{-15} $ & $  <   0.18 \times 10^{-14} $ & $    0.55 $ 	\\
  201 & $  <   26.70 $ & $  <   25.40 $ & $     0.0 $ & $     0.0 $ &   -   & $  <   0.17 \times 10^{-15} $ & $  <   0.96 \times 10^{-15} $ & $    0.53 $ 	\\
  202 & $  <    5.40 $ & $  <   11.70 $ & $     0.0 $ & $     0.0 $ &   -   & $  <   0.14 \times 10^{-15} $ & $  <   0.15 \times 10^{-14} $ & $   - $ 	\\
  203 & $  <    6.70 $ & $  <    4.00 $ & $     0.0 $ & $     0.0 $ &   -   & $  <   0.18 \times 10^{-15} $ & $  <   0.45 \times 10^{-15} $ & $    1.23 $ 	\\
  204 & $  <   11.30 $ & $  <    9.80 $ & $     0.0 $ & $     0.0 $ &   -   & $  <   0.30 \times 10^{-15} $ & $  <   0.14 \times 10^{-14} $ & $    1.13 $ 	\\
  205 & $  <   18.70 $ & $  <   11.20 $ & $     0.0 $ & $     0.0 $ &   -   & $  <   0.51 \times 10^{-15} $ & $  <   0.13 \times 10^{-14} $ & $    0.15 $ 	\\
  209 & $   29.10 \pm    8.60 $ & $ <    44.20 $ & $     3.7 $ & $     0.0 $ & $   -1.00 $ & $   0.40 \times 10^{-15} $ & $ <  0.36 \times 10^{-14} $ & $   - $ 	\\
  210 & $  <    5.30 $ & $  <    8.90 $ & $     0.0 $ & $     0.0 $ &   -   & $  <   0.15 \times 10^{-15} $ & $  <   0.11 \times 10^{-14} $ & $    1.13 $ 	\\
  212 & $  <   23.30 $ & $  <   19.90 $ & $     0.0 $ & $     0.0 $ &   -   & $  <   0.17 \times 10^{-15} $ & $  <   0.85 \times 10^{-15} $ & $    0.58 $ 	\\
  213 & $  <    9.50 $ & $  <   10.90 $ & $     0.0 $ & $     0.0 $ &   -   & $  <   0.24 \times 10^{-15} $ & $  <   0.14 \times 10^{-14} $ & $   - $ 	\\
  216 & $  <    2.90 $ & $  <    7.10 $ & $     0.0 $ & $     0.0 $ &   -   & $  <   0.14 \times 10^{-15} $ & $  <   0.18 \times 10^{-14} $ & $   - $ 	\\
  221 & $  <    9.60 $ & $  <   11.70 $ & $     0.0 $ & $     0.0 $ &   -   & $  <   0.25 \times 10^{-15} $ & $  <   0.16 \times 10^{-14} $ & $    0.69 $ 	\\
  222 & $  <   16.70 $ & $  <   12.30 $ & $     0.0 $ & $     0.0 $ &   -   & $  <   0.46 \times 10^{-15} $ & $  <   0.15 \times 10^{-14} $ & $   - $ 	\\
  223 & $  <   33.80 $ & $  <   33.80 $ & $     0.0 $ & $     0.0 $ &   -   & $  <   0.29 \times 10^{-15} $ & $  <   0.17 \times 10^{-14} $ & $   - $ 	\\
  224 & $  <   38.90 $ & $  <   35.70 $ & $     0.0 $ & $     0.0 $ &   -   & $  <   0.81 \times 10^{-15} $ & $  <   0.45 \times 10^{-14} $ & $    0.56 $ 	\\
  226 & $  <   29.30 $ & $  <   11.20 $ & $     0.0 $ & $     0.0 $ &   -   & $  <   0.34 \times 10^{-15} $ & $  <   0.79 \times 10^{-15} $ & $   - $ 	\\
  227 & $  <    5.60 $ & $  <   19.70 $ & $     0.0 $ & $     0.0 $ &   -   & $  <   0.15 \times 10^{-15} $ & $  <   0.27 \times 10^{-14} $ & $   - $ 	\\
  229 & $   27.70 \pm   10.80 $ & $ <    56.10 $ & $     2.9 $ & $     0.0 $ & $   -1.00 $ & $   0.26 \times 10^{-15} $ & $ <  0.32 \times 10^{-14} $ & $    0.18 $ 	\\
  233 & $  <   12.10 $ & $  <    7.50 $ & $     0.0 $ & $     0.0 $ &   -   & $  <   0.32 \times 10^{-15} $ & $  <   0.10 \times 10^{-14} $ & $    0.53 $ 	\\
  234 & $  <    9.10 $ & $  <    5.60 $ & $     0.0 $ & $     0.0 $ &   -   & $  <   0.25 \times 10^{-15} $ & $  <   0.82 \times 10^{-15} $ & $   - $ 	\\
  235 & $  <   16.30 $ & $  <   22.50 $ & $     0.0 $ & $     0.0 $ &   -   & $  <   0.27 \times 10^{-15} $ & $  <   0.22 \times 10^{-14} $ & $   - $ 	\\
  236 & $  <   26.80 $ & $   30.00 \pm   13.90 $ & $     0.0 $ & $     2.6 $ & $    1.00 $ & $  <   0.28 \times 10^{-15} $ & $   0.19 \times 10^{-14} $ & $   - $	\\
  237 & $  <   17.00 $ & $  <   18.60 $ & $     0.0 $ & $     0.0 $ &   -   & $  <   0.47 \times 10^{-15} $ & $  <   0.28 \times 10^{-14} $ & $   - $ 	\\
  238 & $  <   16.60 $ & $  <   13.90 $ & $     0.0 $ & $     0.0 $ &   -   & $  <   0.45 \times 10^{-15} $ & $  <   0.20 \times 10^{-14} $ & $    1.10 $ 	\\
  239 & $  <   28.60 $ & $  <   37.50 $ & $     0.0 $ & $     0.0 $ &   -   & $  <   0.41 \times 10^{-15} $ & $  <   0.33 \times 10^{-14} $ & $    1.03 $ 	\\
  241 & $  <    3.00 $ & $  <    7.60 $ & $     0.0 $ & $     0.0 $ &   -   & $  <   0.77 \times 10^{-16} $ & $  <   0.10 \times 10^{-14} $ & $    0.57 $ 	\\
  242 & $  <   11.00 $ & $  <    6.00 $ & $     0.0 $ & $     0.0 $ &   -   & $  <   0.28 \times 10^{-15} $ & $  <   0.80 \times 10^{-15} $ & $   - $ 	\\
  244 & $  <   11.10 $ & $  <    9.90 $ & $     0.0 $ & $     0.0 $ &   -   & $  <   0.32 \times 10^{-15} $ & $  <   0.15 \times 10^{-14} $ & $   - $ 	\\
  245 & $  <   12.60 $ & $  <   13.90 $ & $     0.0 $ & $     0.0 $ &   -   & $  <   0.36 \times 10^{-15} $ & $  <   0.22 \times 10^{-14} $ & $   - $ 	\\
  247 & $  <   13.70 $ & $  <   17.90 $ & $     0.0 $ & $     0.0 $ &   -   & $  <   0.38 \times 10^{-15} $ & $  <   0.21 \times 10^{-14} $ & $   - $ 	\\
  248 & $  <   15.20 $ & $  <   22.90 $ & $     0.0 $ & $     0.0 $ &   -   & $  <   0.44 \times 10^{-15} $ & $  <   0.36 \times 10^{-14} $ & $   - $ 	\\
  253 & $  <    7.70 $ & $  <   10.80 $ & $     0.0 $ & $     0.0 $ &   -   & $  <   0.21 \times 10^{-15} $ & $  <   0.13 \times 10^{-14} $ & $    0.54 $ 	\\
  254 & $  <   13.70 $ & $  <   10.00 $ & $     0.0 $ & $     0.0 $ &   -   & $  <   0.38 \times 10^{-15} $ & $  <   0.12 \times 10^{-14} $ & $    0.45 $ 	\\
  255 & $  <   10.80 $ & $  <    4.20 $ & $     0.0 $ & $     0.0 $ &   -   & $  <   0.29 \times 10^{-15} $ & $  <   0.59 \times 10^{-15} $ & $   - $ 	\\
  256 & $  <   10.70 $ & $  <   19.20 $ & $     0.0 $ & $     0.0 $ &   -   & $  <   0.30 \times 10^{-15} $ & $  <   0.23 \times 10^{-14} $ & $    0.86 $ 	\\
  257 & $   11.40 \pm    4.70 $ & $ <    28.00 $ & $     2.5 $ & $     0.0 $ & $   -1.00 $ & $   0.33 \times 10^{-15} $ & $ <  0.35 \times 10^{-14} $ & $    0.56 $ 	\\
  258 & $  <   13.20 $ & $  <   16.50 $ & $     0.0 $ & $     0.0 $ &   -   & $  <   0.39 \times 10^{-15} $ & $  <   0.26 \times 10^{-14} $ & $   - $ 	\\
  260 & $   11.90 \pm    4.70 $ & $ <     9.10 $ & $     2.7 $ & $     0.0 $ & $   -1.00 $ & $   0.34 \times 10^{-15} $ & $ <  0.11 \times 10^{-14} $ & $    0.20 $ 	\\
  261 & $  <   16.60 $ & $  <   12.40 $ & $     0.0 $ & $     0.0 $ &   -   & $  <   0.47 \times 10^{-15} $ & $  <   0.15 \times 10^{-14} $ & $   - $ 	\\
  263 & $  <   17.60 $ & $  <   18.70 $ & $     0.0 $ & $     0.0 $ &   -   & $  <   0.50 \times 10^{-15} $ & $  <   0.23 \times 10^{-14} $ & $    3.68 $ 	\\
  264 & $  <   16.60 $ & $  <   19.00 $ & $     0.0 $ & $     0.0 $ &   -   & $  <   0.47 \times 10^{-15} $ & $  <   0.29 \times 10^{-14} $ & $   - $ 	\\
  265 & $  <    5.20 $ & $  <   11.70 $ & $     0.0 $ & $     0.0 $ &   -   & $  <   0.52 \times 10^{-15} $ & $  <   0.51 \times 10^{-14} $ & $    0.10 $ 	\\
\enddata
\end{deluxetable}

\clearpage

\begin{deluxetable}{rrr}
\tablecaption{ X--ray average luminosity of radio sources with no X--ray counterparts in four redshift bins\label{table5} }
\tablehead{$\langle z \rangle$  & $L_{0.5-2}$ erg s$^{-1}$ &$L_{2-10}$ erg s$^{-1}$}
\startdata  
0.20 &  $ 1.3 \times 10^{40}$  &  $2.1 \times  10^{40}$ \\
0.56 &  $ 9.4 \times 10^{40}$   &  $1.5 \times  10^{41}$ \\
0.78 &  $5.9 \times  10^{40} $   &  $9.7 \times  10^{40}$ \\
1.40 &  $5.8 \times 10^{41}$  &  $1.2 \times  10^{42}$  \\
\enddata
\end{deluxetable}


\begin{thebibliography}{}
\bibitem[]{} Afonso, J., Mobasher, B., Koekemoer, A., Norris, R.P., Cram, 
L. 2006, AJ, 131, 1216
\bibitem[]{} Alexander, D.M., et al. 2003, AJ, 126, 539
\bibitem[]{} Antonucci, R.R.J. 1993, ARA\&A, 31, 473
\bibitem[]{} Arnaud, K.A. 1996, ``Astronomical Data Analysis Software
and Systems V'', eds. Jacoby G. and Barnes J., ASP Conf. Series
vol. 101, 17 
\bibitem[]{} Barger, A.J., Cowie, L.L., \& Wang, W.-H. 2007, ApJ, 654, 764
\bibitem[]{} Bauer, F.E., Alexander, D.M., Brandt, W.N., Hornschemeier, 
A.E., Vignali, C., Garmire, G.P., \& Schneider, D.P. 2002, AJ, 124, 2351 
\bibitem[]{} Bauer, F.E., Alexander, D.M., Brandt, W.N., Schneider, D.P., 
Treister, E., Hornschemeier, A.E., \& Garmire, G.P. 2004, AJ, 128, 2048
\bibitem[]{} Bautz, M., et al. 1998, in Proc. SPIE Vol. 3444, X--ray Optics, 
Instruments and Missions, eds. R.B. Hoover \& A.B. Walker, 210 
\bibitem[]{}Dickey \& Lockman, 1990, ARAA, 28, 215. 
\bibitem[]{}Garmire, G.P., et al. 1992, ApJ, 399, 694
\bibitem[]{} Giacconi, R., Rosati, P., Tozzi, P., et al. 2001, ApJ,
551, 624
\bibitem[]{} Giacconi, R., Zirm, A., Wang, J., Rosati, P., Nonino, M.,
Tozzi, P., Gilli, R., Mainieri, V., Hasinger, G., Kewley, L., et
al. 2002, ApJS, 139, 369
\bibitem[]{} Giavalisco et al. 2004, ApJL, 600, 93
\bibitem[]{} Gilli, R., Cimatti, A., Daddi, E., Hasinger, G., 
Rosati, P., Szokoly, G., Tozzi, P., Bergeron, J., Borgani, S., 
Giacconi, R., Kewley, L., Mainieri, V., Mignoli, M., Nonino, M., Norman, C., 
Wang, J., Zamorani, G., Zheng, W., \& Zirm, A. 2003, ApJ, 592, 721
\bibitem[]{} Hildebrandt et al. 2006, A\&A 452, 1121-1128
\bibitem[]{} Kellermann, K.I., Sramek, R., Schmidt, M., Shaffer, D.B., \& 
R. Green 1989, AJ, 98, 1195
\bibitem[]{} Kellermann, K.I, et al. 2008, ApJS, 179, 71 (Paper I)
\bibitem[]{}Kormendy, J., Richstone, D. 1995, ARA\&A, 33, 581
\bibitem[]{}Lehmer, B. D., Brandt, W. N., Alexander, D. M., Bauer, F. E.,
Schneider, D. P.,  Tozzi, P., Bergeron, J.,  Garmire, G. P., Giacconi, R., 
Gilli, R.,  Hasinger, G., Hornschemeier, A. E.,  Koekemoer, A. M.,
Mainieri, V., Miyaji, T.,  Nonino, M.,  Rosati, P., Silverman, J. D.,
Szokoly, G. \&  Vignali C. 2005, ApJS, 161, 21
\bibitem[]{}Lehmer, B. D., Brandt, W. N., Alexander, D. M., Bell, E. F., 
Hornschemeier, A. E., McIntosh, D. H., Bauer, F. E., Gilli, R., Mainieri, V., 
Schneider, D. P., Silverman, J. D., Steffen, A. T., Tozzi, P., \& Wolf, C. 
2008, ApJ, 681, 1163
\bibitem[]{} Magorrian, J., Tremaine, S., et al. 1998, AJ, 115, 2285
\bibitem[]{} Mainieri, V., Rosati, P., Tozzi, P., Bergeron, J., Gilli, R., 
Hasinger, G., Nonino, M., Idzi, R., Koekemoer, A.M., Lehmann, I., Szokoly, 
G., \& Zheng, W. 2005, A\&A, 437, 805
\bibitem[]{} Mainieri, V., et al. 2008, ApJS, 179, 95 (Paper II)
\bibitem[]{}Muxlow, T. W. B., Richards, A. M. S., Garrington, S. T., 
Wilkinson, P. N., Anderson, B., Richards, E. A., Axon, D. J., 
Fomalont, E. B., Kellermann, K. I., Partridge, R. B., 
Windhorst, R. A. 2005, MNRAS, 358, 1159
\bibitem[]{} Nandra, K., Pounds, K.A. 1994, MNRAS, 268, 405
\bibitem[]{}Panessa, F., Barcons, X., Bassani, L., Cappi, M., Carrera, F.J., 
Ho, L.C., \& Pellegrini, S. 2007, A\&A, 467, 519
\bibitem[]{} Padovani, P., et al. 2009, ApJ, in press, arXiv:0812.2997 
(Paper IV)
\bibitem[]{}Paolillo, M., Schreier, E. J., Giacconi, R., Koekemoer, A.M., \&
Grogin, N. A. 2004, ApJ, 611, 93 
\bibitem[]{} Persic, M., Rephaely, Y.  2007, A\&A, 463, 481
\bibitem[]{} Ranalli, P., Comastri, A., \& Setti, G. 2003, A\&A, 399, 39
\bibitem[]{}Richards, E.A., Kellermann, K.I., Fomalont, E.B., Windhorst, R.A.,
\& Partridge R.B. 1998, AJ, 116, 1039
\bibitem[]{}Richards, E.A. 2000, ApJ, 533, 611
\bibitem[]{}  Rix H.--W. et al. 2004, ApJS, 152, 163
\bibitem[]{} Rovilos, E., Georgakakis, A., Georgantopoulos, I., Afonso, J., 
Koekemoer, A.M., Mobasher, B., \& Goudis, C. 2007, A\&A, 466, 119
\bibitem[]{} Rosati, P., Tozzi, P., Giacconi, R., Gilli, R., Hasinger, G.,
Kewley, L., Mainieri, V., Nonino, M., Norman, C., Szokoly, G., and 9 
coauthors 2002, ApJ, 566, 667
\bibitem[]{} Schinnerer, E., Smolcic, V., Carilli C.L., et al. 2007, 
ApJS, 172, 46 
\bibitem[]{}Silverman et al. 2008, in preparation
\bibitem[]{}Spergel, D.N., Bean, R., Dore', O., Nolta, M.R., et al. 2007, 
ApJS, 170, 377
\bibitem[]{} Smolcic, V., Schinnerer, E., Scodeggio, M., Franzetti, P., et al.
2008a, ApJ in press, arXiv:0803.0997
\bibitem[]{} Smolcic, V., Schinnerer, E., Zamorani, G., et al. 2008b, ApJ 
in press, arXiv:0808.0493
\bibitem[]{} Szokoly, G. P., Bergeron, J., Hasinger, G., Lehmann, I., 
Kewley,  L., Mainieri, V., Nonino, M., Rosati, P., Giacconi, R.,
Gilli, R., Gilmozzi, R., Norman, C., Romaniello, M., Schreier, E., Tozzi, P., 
Wang, J.X., Zheng, W., Zirm, A. 2004, ApJS, 155, 271
\bibitem[]{}Terashima, Y., \& Wilson, A.S. 2003, ApJ, 583, 145
\bibitem[]{}Tozzi, P., Rosati, P., Nonino, M., Bergeron, J., Borgani,
S., Gilli, R., Gilmozzi, R., Hasinger, G., Grogin, N., Kewley, L., et
al. 2001, ApJ, 562, 42
\bibitem[]{} Tozzi, P., Gilli, R., Mainieri, V., Norman, C., Risaliti, G., 
Rosati, P., Bergeron, Giacconi, R., J., Hasinger, G., Nonino, M., 
Streblyanska, A., Szokloly, G., Wang, J.X., \& Zheng, W. 2006, A\&A, 451, 457
\bibitem[]{} Veilleux, S. ,\& Osterbrock, D.E. 1987, ApJS, 63, 295
\bibitem[]{} Vikhlinin, A., Markevitch, M., Murray, S.S., Jones, C., Forman, 
W., \& Van Speybroeck, L. 2005, ApJ, 628, 655
\bibitem[]{}Zheng, W., Mikles, V.J., Mainieri, V., Hasinger, G., 
Rosati, P., Wolf, C., Norman, C., Szokoly, G., Gilli, R., Tozzi, P., Wang, 
J.X., Zirm, A., \& Giacconi, R. 2004, ApJS, 155, 73
\end{thebibliography}
\end{document}